\documentclass{statsoc}

\usepackage{geometry}
\usepackage{anyfontsize}
\usepackage{color}

\usepackage[utf8]{inputenc}

\usepackage{array}
\usepackage{amsmath}
\usepackage{amssymb}
\usepackage{bbm}
\usepackage{mathtools}
\usepackage{graphicx}
\usepackage{siunitx}
\usepackage{pgf,tikz,pgfplots}
\pgfplotsset{compat=1.15}
\usepackage{mathrsfs}
\usepackage{transparent}
\usetikzlibrary{arrows}
\usetikzlibrary{calc}

\usepackage{etoolbox}
\makeatletter
\patchcmd{\@makecaption}
{\parbox}
{\advance\@tempdima-\fontdimen2} 
{}{}
\makeatother  

\usepackage{zref-savepos}
\usepackage{tabu}

\newcounter{NoTableEntry}
\renewcommand*{\theNoTableEntry}{NTE-\the\value{NoTableEntry}}

\newcommand*{\strike}[2]{%
	\multicolumn{1}{#1}{%
		\stepcounter{NoTableEntry}%
		\vadjust pre{\zsavepos{\theNoTableEntry t}}
		\vadjust{\zsavepos{\theNoTableEntry b}}
		\zsavepos{\theNoTableEntry l}
		\hspace{0pt plus 1filll}%
		#2
		\hspace{0pt plus 1filll}%
		\zsavepos{\theNoTableEntry r}
		\tikz[overlay]{%
			\draw
			let
			\n{llx}={\zposx{\theNoTableEntry l}sp-\zposx{\theNoTableEntry r}sp-\tabcolsep},
			\n{urx}={\tabcolsep},
			\n{lly}={\zposy{\theNoTableEntry b}sp-\zposy{\theNoTableEntry r}sp},
			\n{ury}={\zposy{\theNoTableEntry t}sp-\zposy{\theNoTableEntry r}sp}
			in
			(\n{llx}, \n{lly}) -- (\n{urx}, \n{ury})
			;
		}%
	}%
}

\usepackage[normalem]{ulem}
\usepackage{array}
\newcolumntype{P}[1]{>{\centering\arraybackslash}p{#1}}
\newcolumntype{M}[1]{>{\centering\arraybackslash}m{#1}}

\newcommand*\samethanks[1][\value{footnote}]{\footnotemark[#1]}
\usepackage{comment}

\usepackage{hyperref}

\usepackage{natbib}
\newtheorem{prop}{Proposition}

\date{\today}
\title[Saturated pairwise interaction Gibbs point process]{The saturated pairwise interaction Gibbs point process as a joint species distribution model}
\author{Ian Flint\thanks{Corresponding author: {\sf{ian.flint@unimelb.edu.au}}.}\thanks{Work supported by Australian Research Council Grant No DP190100613.},}
\address{School of Ecosystem  and Forest Sciences, the University of Melbourne, Parkville VIC, Australia}
\author{Nick Golding\samethanks,}
\address{School of Public Health, Curtin University, Perth WA, Australia}
\author{Peter Vesk\samethanks,}
\address{School of Ecosystem  and Forest Sciences, the University of Melbourne, Parkville VIC, Australia}
\author{Yan Wang\samethanks}
\address{School of Science, RMIT University, 124 La Trobe St, Melbourne VIC, Australia}
\author[Ian Flint {\it et al.}]{and Aihua Xia\samethanks}
\address{School of Mathematics and Statistics, the University of Melbourne, Parkville VIC, Australia}

\begin{document}
	\maketitle
	
	\begin{abstract}
		In an effort to effectively model observed patterns in the spatial configuration of individuals of multiple species in nature, we introduce the saturated pairwise interaction Gibbs point process.
		Its main strength lies in its ability to model both attraction and repulsion within and between species, over different scales.
		As such, it is particularly well-suited to the study of associations in complex ecosystems.
		Based on the existing literature, we provide an easy to implement fitting procedure as well as a technique to make inference for the model parameters.
		We also prove that under certain hypotheses the point process is locally stable, which allows us to use the well-known `coupling from the past' algorithm to draw samples from the model.
		Different numerical experiments show the robustness of the model.
		We study three different ecological datasets, demonstrating in each one that our model helps disentangle competing ecological effects on species' distribution.
	\end{abstract}
	
	
	\section{Introduction}
	
	Point processes, i.e., random events in time and/or space, have seen widespread use in forestry and plant ecology \citep{TH55}, astronomy \citep{BF96}, epidemiology \citep{WG04}, geology \citep{CH95}, wireless networks \citep{A10,BB1} and criminology \citep{MSBST}.
	A marked point process is a point process that has additional features attached to each event.
	Marked point processes are particularly important in ecology where observations often include properties of the individuals, for example their size or species.
	Indeed, marked point processes have been used to model trees and their species in a swamp forest \citep{D}, trees in a tropical forest along with their species and size \citep{HCF}, adult and juvenile plants in a dipterocarp forest along with their species \citep{PGWKGGWH}, see also the many other examples in the \verb|spatstat.data| R package \citep{spatstat}.
	
	The spatial arrangement of individuals of different species reflects what is termed ecological community assembly.
	Community assembly may be thought as the outcome of species having differential dispersal abilities, environmental tolerance and biotic interactions \citep{weiher2011advances}.
	Statistical modelling of community assembly data on multi-species abundances or presence/absence in samples has focused on environmental tolerance and biotic interactions, within the framework of generalised linear mixed effects modelling \citep{ovaskainen2017make}.
	Yet, community assembly is an outcome of underlying mechanisms of birth, growth, reproduction, dispersal and death of individuals. 
	Those individuals' fates may be affected by other individuals of the same or different species through positive and negative interactions, such as competition and facilitation.
	Additionally, indirect interactions between individuals within a species may be mediated through natural enemies (e.g., seed predators, herbivores and pathogens) that use density-dependent search strategies.
	Compared to alternative models that work with abundance or occurrence data, the use of point process models in such ecological settings enables inferences about the underlying mechanisms driving spatial arrangement of individuals in a multi-species setting. 
	
	The most common approach to the analysis of multi-species point processes is to compute the cross-pair correlation functions between all pairs of species, following the methodology of \citet{MW} (see Sections~4.4 and 4.5 therein).
	Such an approach quickly becomes impractical when the number of species increases beyond the bivariate setting.
	Compared to this type of ad-hoc analysis, an integrated modelling framework is more robust and allows ecological questions to be answered more systematically.
	
	A number of such models have recently been introduced to tackle multi-species spatial point patterns.
	First, the log-Gaussian Cox process has been successfully used to jointly model the locations of a nine tree species subset of the Barro Colorado Island 50 Ha plot in \cite{WGJM}.
	Briefly, the model assumes that a number of correlated Gaussian fields are driving the log-intensity of the point process, and the correlation coefficients between the Gaussian fields are thought to represent positive or negative interactions between species. 
	Second, some types of Gibbs point processes have been used to model a larger subset of species from the same Barro Colorado Island dataset in \cite{RMO}.
	Although the Gibbs point process is usually thought of as modelling repulsion, the Geyer model~\citep{G} introduces a saturation parameter that allows it to model either attraction or repulsion.
	The model in \cite{RMO} is an extended version of the Geyer model adapted to the multi-species setting.
	
	Our aim in this manuscript is to expand upon some of the ideas in \cite{RMO} and to build a solid unified framework that can be applied to a wide range of multi-species marked point patterns.
	To that end, we introduce the saturated pairwise interaction Gibbs point process.
	Within this class of models, \cite{RMO} only consider potential functions that are linear combinations of step functions and in brief, our model instead is based on ecologically sound potential functions.
	We also allow for pairwise interactions whose magnitude are driven by the individuals' marks, such as their size, which are thought to be important in explaining species' distribution.
	
	In addition to providing the theoretical background to the model, we shall also consider the problem of simulating the point process without conditioning on the number of points.
	Although not considered in \cite{RMO}, such unconditioned simulation is indeed important, e.g., to do Monte-Carlo simulations as well as compute simulation envelopes and run goodness of fit tests.
	The model is validated in a series of numerical experiments in which we compute coverage probabilities, mean estimates, and consider the sensitivity of the inference procedure to some of the fixed model parameters.
	
	We demonstrate the flexibility of our model by applying it to various point patterns of interest to plant ecologists.
	These cover a range of ecosystems and, within the framework of our model, showcase the different types of positive and negative interactions that arise in the analysis of ecological data.
	We start with modelling the locations of a hundred Norway spruce trees in Germany \citep{F} before moving on to a study of close to a thousand trees in a swamp forest in South Carolina \citep{GW}. 
	We conclude with an analysis of the well-known Barro Colorado Island dataset studied in \cite{RMO} and \cite{WGJM}.
	In this last analysis, we cover a larger subset of the data than \cite{WGJM}, and compared to \cite{RMO}, we put more emphasis on the species' interactions.
	We examine how well our model has performed on a dataset containing almost a hundred species and many thousands of individual trees.
	In each of these examples, we show that our model has helped quantify and disentangle the effects of different ecological mechanisms on the spatial distribution.
	The model described in this manuscript has been implemented as an R package (R Core Team, 2019) to facilitate its use by ecology practitioners and other researchers.
	
	We begin in Section~\ref{sec:specification} by introducing some notation and defining our new Gibbs point process.
	In Section~\ref{sec:inference} we explain how to do inference on our point process model by using the logistic regression inference technique for Gibbs point processes from \cite{BCRW}.
	We recall two important simulation algorithms in Section~\ref{sec:simulation}, and prove that they can be applied to our model.
	We provide numerical studies in Section~\ref{sec:numerical} and applications to real datasets in Section~\ref{sec:datasets}.
	Some additional technical results are given in the supplementary material.
	
	\section{Mathematical specification of the model}
	\label{sec:specification}
	
	\subsection{Notation}
	
	Before introducing our model, we start with some brief elements of point process theory.
	The interested reader may find more details in any of the numerous textbooks on the topic, for example \cite{K}, \cite{DV}, \cite{DV2} and \cite{MW}.
	In the remainder of this manuscript, we shall often use $\mathbf1_A$ to denote the indicator function of a set $A$, i.e., the function with values $1$ on $A$ and $0$ elsewhere.
	
	We consider $p$ species located in a bounded region $W\subset\mathbb{R}^2$, with individuals labeled by an integer representing their species, and each individual also equipped with a mark in $\mathbb{R}$ representing one of its properties such as its height or width. 
	More precisely, we denote by $(x,i,m)\in W\times\{1,\dots, p\}\times\mathbb{R}=:\mathbb{S}$ an individual of species $i$ at the location $x$ with a mark $m$. 
	The set of admissible configurations is denoted by
	\begin{equation*}
	\mathcal{N}:=\bigl\{\omega\;:\;\omega\subset\mathbb{S},\ \omega\mbox{ is a finite set}\bigr\},
	\end{equation*}
	and the cardinality of the set $\omega$ by $|\omega|$.
	We assume that the reference measure of the marks is the Lebesgue measure.
	Given a configuration of individuals $\omega\in\mathcal{N}$, we denote by $\omega_i$ the sub-configuration consisting of individuals of species $i$, for $1\le i\le p$.
	In short, $\omega_i:=\{(x,k,m)\in\omega\;:\;k=i\}$.
	
	In point process theory, a number of descriptive functions have been introduced.
	The two we shall use in this manuscript are the Papangelou conditional intensity and the density with respect to a unit rate Poisson point process.
	Briefly, the density of the point process is the function $j$ such that $j(\omega)$ is proportional to the probability that a configuration occurs in an infinitesimal volume around $\omega\in\mathcal N$.
	
	The Papangelou conditional intensity of the point process~\cite[Definition~10.4.I]{DV2} is a function $\pi$ such that $\pi((x,i,m),\omega)\,\mathrm dx\mathrm dm$ is the probability that there is an individual of species $i$, mark around $m$ and location around $x\in W$ conditional on the configuration $\omega\in\mathcal{N}$ (outside of $\mathrm dx\times\{i\}\times\mathrm dm$).
	Contrary to the density $j$, the Papangelou conditional intensity gives information on the conditional probability of finding new individuals, given an existing configuration.
	Although the density characterises the point process, it is the Papangelou conditional intensity that appears in point process inference and simulation, and thus it shall play a key role in our analysis.
	
	\subsection{Potential functions}
	\label{subsec:potential}
	
	Our model is based on short- and medium-range potential functions, which themselves depend on interaction distances.
	In order to facilitate comparisons between different potential functions, we impose a few conditions.
	A \textit{short-range potential function} $\varphi_{R^{\mathrm S}}$ with short-range interaction radius $R^{\mathrm S}$ is a $[0,1]$-valued decreasing$^{\footnotemark}$
	\footnotetext{Hereon, the term `decreasing' is to be understood in the weak sense, i.e., $\varphi$ is said to be decreasing if for all $x\le y$, $\varphi(x)\ge\varphi(y)$.} function which satisfies $\varphi_{R^{\mathrm S}}(0)=1$, $\varphi_{R^{\mathrm S}}(r)\ge0.5$ for $r\le R^{\mathrm S}$ and $\varphi_{R^{\mathrm S}}(r)<0.5$ for $r>R^{\mathrm S}$.
	Similarly, a \textit{medium-range potential function} $\psi_{R^{\mathrm M}\leftrightarrow R^{\mathrm L}}$ with medium-range interaction radius $R^{\mathrm M}$ and long-range interaction radius $R^{\mathrm L}$ is a $[0,1]$-valued function such that $0.5\le\psi_{R^{\mathrm M}\leftrightarrow R^{\mathrm L}}(r)\le1$ for $R^{\mathrm M}\le r\le R^{\mathrm L}$ and $0\le\psi_{R^{\mathrm M}\leftrightarrow R^{\mathrm L}}(r)<0.5$ otherwise.
	Larger values of the potential functions correspond to stronger interactions, and in particular the potential functions fall to no more than half of their maximum at the corresponding interaction distances.
	We list in Table~\ref{tab:potentials} a few commonly used potential functions at short and medium ranges.

	\begin{table}
		\caption{\label{tab:potentials} Short-range (first three rows, in blue) and medium-range (last two rows, in red) potential functions.
		In the table below, $R^{\mathrm S}$, $R^{\mathrm M}$ and $R^{\mathrm L}$ are, respectively, the short-range, medium-range and long-range interaction radii.}
		\scalebox{0.7}{
		\begin{tabular}{|>{\centering\arraybackslash}m{.14\linewidth}|>{\centering\arraybackslash}m{.32\linewidth}|>{\centering\arraybackslash}m{.35\linewidth}|>{\centering\arraybackslash}m{.1\linewidth}|}
			\hline
			Potential function & Definition & Shape\\
			\hline\hline
			Exponential & $\exp\bigl[-r\,\ln(2)/R^{\mathrm S}\bigr]$ & \begin{tikzpicture}[x=1cm,y=1cm,scale=1]
			\begin{axis}[
			x=1cm,
			y=1cm,
			line width=1pt,
			axis lines=middle,
			xmajorgrids=true,
			xminorgrids=true,
			xmin=0,
			xmax=3.5,
			ymajorgrids=true,
			yminorgrids=true,
			ymin=0,
			ymax=1.5,
			xtick={0,1,2,3},
			ytick={0,1},
			minor tick num=1]
			\clip(0,0) rectangle (3.5,1.5);
			\end{axis}
			\draw (-0.1,-0.25) node {$0$};
			\draw[color=blue] (0.5,-0.3) node {\scriptsize$R^{\mathrm S}=$};
			\draw[line width=1pt,scale=1,domain=0:3.5,smooth,samples=100,variable=\x,opacity=1,color=blue] plot ({\x},{exp(-\x * ln(2) / 1)});
			\end{tikzpicture}\\
			\hline
			Square bump & $1-\exp\bigl[-(R^{\mathrm S})^2\,\ln(2)/r^2\bigr]$ & \begin{tikzpicture}[x=1cm,y=1cm,scale=1]
			\begin{axis}[
			x=1cm,
			y=1cm,
			line width=1pt,
			axis lines=middle,
			xmajorgrids=true,
			xminorgrids=true,
			xmin=0,
			xmax=3.5,
			ymajorgrids=true,
			yminorgrids=true,
			ymin=0,
			ymax=1.5,
			xtick={0,1,2,3},
			ytick={0,1},
			minor tick num=1]
			\clip(0,0) rectangle (3.5,1.5);
			\end{axis}
			\draw (-0.1,-0.25) node {$0$};
			\draw[color=blue] (0.5,-0.3) node {\scriptsize$R^{\mathrm S}=$};
			\draw[line width=1pt,scale=1,domain=0.01:3.5,smooth,samples=100,variable=\x,opacity=1,color=blue] plot ({\x},{1 - exp(-1 * 1 * ln(2) / (\x * \x))});
			\end{tikzpicture}\\
			\hline
			Step & $\mathbf1_{[0,R^{\mathrm S}]}(r)$ & \begin{tikzpicture}[x=1cm,y=1cm,scale=1]
			\begin{axis}[
			x=1cm,
			y=1cm,
			line width=1pt,
			axis lines=middle,
			xmajorgrids=true,
			xminorgrids=true,
			xmin=0,
			xmax=3.5,
			ymajorgrids=true,
			yminorgrids=true,
			ymin=0,
			ymax=1.5,
			xtick={0,1,2,3},
			ytick={0,1},
			minor tick num=1]
			\clip(0,0) rectangle (3.5,1.5);
			\end{axis}
			\draw (-0.1,-0.25) node {$0$};
			\draw[color=blue] (0.5,-0.3) node {\scriptsize$R^{\mathrm S}=$};
			\draw[line width=1pt,scale=1,opacity=1,color=blue] (0,1) -- (1,1);
			\draw[line width=1pt,scale=1,opacity=1,color=blue] (1,1) -- (1,0);
			\draw[line width=1pt,scale=1,opacity=1,color=blue] (1,0) -- (3.3,0);
			\end{tikzpicture}\\
			\hline\hline
			Normal & $\exp\biggl[-\frac{4(r-(R^{\mathrm M}+R^{\mathrm L})/2)^2\,\ln(2)}{(R^{\mathrm L}-R^{\mathrm M})^2}\biggr]$ & \begin{tikzpicture}[x=1cm,y=1cm,scale=1]
			\begin{axis}[
			x=1cm,
			y=1cm,
			line width=1pt,
			axis lines=middle,
			xmajorgrids=true,
			xminorgrids=true,
			xmin=0,
			xmax=3.5,
			ymajorgrids=true,
			yminorgrids=true,
			ymin=0,
			ymax=1.5,
			xtick={0,1,2,3},
			ytick={0,1},
			minor tick num=1]
			\clip(0,0) rectangle (3.5,1.5);
			\end{axis}
			\draw (-0.1,-0.25) node {$0$};
			\draw[color=red] (0.5,-0.3) node {\scriptsize$R^{\mathrm M}=$};
			\draw[color=red] (1.5,-0.3) node {\scriptsize$R^{\mathrm L}=$};
			\draw[line width=1pt,scale=1,domain=0:3.5,smooth,samples=100,variable=\x,opacity=1,color=red] plot ({\x},{exp(-4 * (\x-(1+2)/2) * (\x-(1+2)/2) * ln(2) / ((1-2) * (1-2)))});
			\end{tikzpicture}\\
			\hline
			Geyer & $\mathbf1_{[R^{\mathrm M},R^{\mathrm L}]}(r)$ & \begin{tikzpicture}[x=1cm,y=1cm,scale=1]
			\begin{axis}[
			x=1cm,
			y=1cm,
			line width=1pt,
			axis lines=middle,
			xmajorgrids=true,
			xminorgrids=true,
			xmin=0,
			xmax=3.5,
			ymajorgrids=true,
			yminorgrids=true,
			ymin=0,
			ymax=1.5,
			xtick={0,1,2,3},
			ytick={0,1},
			minor tick num=1]
			\clip(0,0) rectangle (3.5,1.5);
			\end{axis}
			\draw (-0.1,-0.25) node {$0$};
			\draw[color=red] (0.5,-0.3) node {\scriptsize$R^{\mathrm M}=$};
			\draw[color=red] (1.5,-0.3) node {\scriptsize$R^{\mathrm L}=$};
			\draw[line width=1pt,scale=1,opacity=1,color=red] (0,0) -- (1,0);
			\draw[line width=1pt,scale=1,opacity=1,color=red] (1,0) -- (1,1);
			\draw[line width=1pt,scale=1,opacity=1,color=red] (1,1) -- (2,1);
			\draw[line width=1pt,scale=1,opacity=1,color=red] (2,1) -- (2,0);
			\draw[line width=1pt,scale=1,opacity=1,color=red] (2,0) -- (3.3,0);
			\end{tikzpicture}\\
			\hline
		\end{tabular}}
	\end{table} 
	
	\subsection{Model}
	
	Our model is parametrised by the following quantities.
	\begin{enumerate}
		\item An intercept vector $(\beta_{1,0},\ldots,\beta_{p,0})^T\in\mathbb{R}^p$ which is interpreted as the log-intensities of the different species, if there were no interactions.
		\item Environmental covariates $X_1,\dots,X_K$ which are assumed to be bounded.
		\item For $1\le i\le p$ and $1\le k\le K$, a coefficient $\beta_{i,k}$ that represents the response of species $i$ to environmental covariate $k$.
		\item\label{ie:d} A function $u(z,(\omega\setminus\{z\})_{i_2})$ representing the short-range interactions between species $i_2$ in $\omega$ and an individual $z=(x,i_1,m)$ of species $i_1$ with mark $m$ at location $x$.
		\item A function $v(z,(\omega\setminus\{z\})_{i_2})$ that models the medium-range interactions between species $i_2$ in $\omega$ and an individual $z$ as in \eqref{ie:d}.
		\item\label{it:alpha} For $1\le i_1,i_2\le p$, a coefficient $\alpha_{i_1,i_2}$ which represents the magnitude of short-range interactions between species $i_1$ and species $i_2$. 
		Positive values of $\alpha_{i_1,i_2}$ correspond to attraction between species $i_1$ and species $i_2$ while negative values are associated with repulsion.
		Note that it is assumed that $\alpha$ is \textit{symmetric}, in the sense that $\alpha_{i_1,i_2}=\alpha_{i_2,i_1}$.
		\item For $1\le i_1,i_2\le p$, a symmetric coefficient $\gamma_{i_1,i_2}$ which is the magnitude of medium-range interactions between each pair of species $i_1$ and species $i_2$.
		As in \eqref{it:alpha}, we interpret the sign of $\gamma_{i_1,i_2}$ as indicating either attraction or repulsion.
	\end{enumerate}
	The model is specified by its density, defined by
	\begin{multline}
	\label{eq:ModelJanossy}
	j(\omega)
	=C\exp\Biggl[\sum_{(x,i,m)\in\omega}\biggl(\beta_{i,0}+\sum_{k=1}^K\beta_{i,k}X_k(x)\biggr)\\
	+\sum_{i_2=1}^p\sum_{z=(x_1,i_1,m_1)\in\omega}\alpha_{i_1,i_2}u(z,(\omega\setminus\{z\})_{i_2})
	+\sum_{i_2=1}^p\sum_{z=(x_1,i_1,m_1)\in\omega}\gamma_{i_1,i_2}v(z,(\omega\setminus\{z\})_{i_2})\Biggr],
	\end{multline}
	for $\omega\in\mathcal{N}$, and where $C>0$ is a normalising constant.
	The Papangelou conditional intensity $\pi$ directly follows from \eqref{eq:ModelJanossy} by the formula $\pi((x,i,m),\omega):=j(\omega\cup\{(x,i,m)\})/j(\omega)$ for $(x,i,m)\notin\omega$.
	We compute $\pi$ explicitly in the supplementary material.
	
	As mentioned above, the function $u(z,(\omega\setminus\{z\})_{i_2})$ is interpreted as the saturated sum of short-range interactions between species $i_2$ in $\omega$ and an individual $z=(x_1,i_1,m_1)$ of species $i_1$ at $x_1$ and with mark $m_1$.
	Letting $R^{\mathrm S}_{i_1,i_2}$ denote the short range interaction distance between species $i_1$ and $i_2$, we propose to define $u$ as either
	\begin{equation}
	\label{eq:unMarkedU}
		u_{\text{unmarked}}((x_1,i_1,m_1),\omega_{i_2})
		:=\max_{\eta\in S(\omega_{i_2},N)}\sum_{(x_2,i_2,m_2)\in\eta}\varphi_{R^{\mathrm S}_{i_1,i_2}}(\|x_1-x_2\|),
	\end{equation}
	or, taking into account marks,
	\begin{equation}
	\label{eq:markedU}
		u_{\text{marked}}((x_1,i_1,m_1),\omega_{i_2})
		:=\max_{\eta\in S(\omega_{i_2},N)}\sum_{(x_2,i_2,m_2)\in\eta}\varphi_{R^{\mathrm S}_{i_1,i_2}}\biggl(\frac{2\,\|x_1-x_2\|}{m_1+m_2}\biggr),
	\end{equation}
	where $N$ is called the saturation parameter, and the set of saturated configurations is defined as $S(\omega,N)=\{\eta\subset\omega\;:\;|\eta|\le N\}$.
	The quantity $u_{\text{unmarked}}((x_1,i_1,m_1),\omega_{i_2})$ consists in the sum of the $N$ largest pairwise interactions between the individual at $x_1$ and individuals of species $i_2$ in $\omega$.
	Heuristically, the larger this quantity, the more short-range interactions there are between the individual at $x_1$ and species $i_2$.
	Our interpretation of the saturation parameter $N$ is similar to that of \cite{RMO} who write that $N$ ``reproduces the feature that the neighbourhood must eventually saturate with individuals as resources are finite''.
	
	In the first of our two definitions \eqref{eq:unMarkedU}, the distances $R^{\mathrm S}_{i_1,i_2}$ are interpreted as typical short-range interaction distances between individuals of species $i_1$ and $i_2$.
	This contrasts with the second definition \eqref{eq:markedU}, in which the distances $R^{\mathrm S}_{i_1,i_2}$ are measured as a proportion of the average marks of interacting individuals.
	One could consider other choices involving marks instead of \eqref{eq:markedU}, for example interactions proportional to the absolute difference of marks, thereby modelling fiercer competition between dissimilar individuals.
	
	Similarly, letting $R^{\mathrm M}_{i_1,i_2}$ (respectively $R^{\mathrm L}_{i_1,i_2}$) be the medium- (respectively long-) range interaction distances between species $i_1$ and $i_2$, we define
	\begin{equation}
	\label{eq:unMarkedV}
	v_{\text{unmarked}}((x_1,i_1,m_1),\omega_{i_2})
	:=\max_{\eta\in S(\omega_{i_2},N)}\sum_{(x_2,i_2,m_2)\in\eta}\psi_{R^{\mathrm M}_{i_1,i_2}\leftrightarrow R^{\mathrm L}_{i_1,i_2}}(\|x_1-x_2\|),
	\end{equation}
	as well as
	\begin{equation}
	\label{eq:markedV}
	v_{\text{marked}}((x_1,i_1,m_1),\omega_{i_2})
	:=\max_{\eta\in S(\omega_{i_2},N)}\sum_{(x_2,i_2,m_2)\in\eta}\psi_{R^{\mathrm M}_{i_1,i_2}\leftrightarrow R^{\mathrm L}_{i_1,i_2}}\biggl(\frac{2\,\|x_1-x_2\|}{m_1+m_2}\biggr),
	\end{equation}
	where the set of saturated configurations $S(\omega_{i_2},N)$ was defined above.
	The parameters $R^{\mathrm M}_{i_1,i_2}$ and $R^{\mathrm L}_{i_1,i_2}$ have the same interpretation as $R^{\mathrm S}_{i_1,i_2}$, but relate to what we call medium- and long-range interactions instead of short-range ones.
	
	\subsection{Saturated pairwise interaction Gibbs point process}
	
	We call our model defined by \eqref{eq:ModelJanossy} a `saturated pairwise interaction Gibbs point process', and the aim of this section is to make explicit why we have settled on this name. 
	As an aside, although to the best of our knowledge saturated pairwise interaction Gibbs point processes have never been described in the scientific literature, \verb|spatstat| has implemented internally what they call \verb|pairsat.family| and describe as a ``Saturated Pairwise Interaction Point Process Family''.
	
\begin{figure}[!ht]
		\centering
		\includegraphics[scale=0.3]{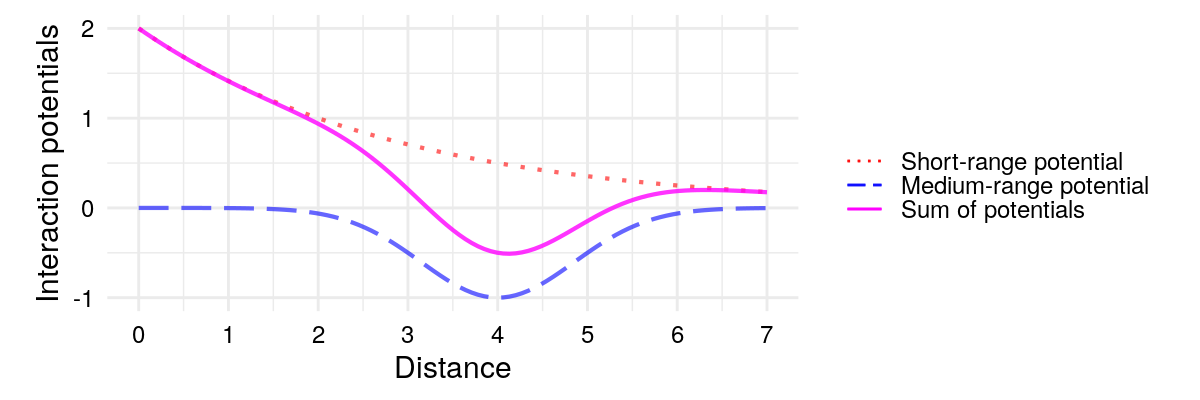}
		\caption{\label{fig:sum_potentials} Two potential functions summed together, for $\alpha=1$, $\gamma=-1/2$, exponential short-range potential, and normal medium-range potential (see Table~\ref{tab:potentials}), when $N=\infty$.
			We plot the short-range potential in densely dashed red (-\,-\,-\,-\,-), the medium-range potential in loosely dashed blue ($-\,-\,-$), and the sum of the two in solid purple (\rule[0.5ex]{2em}{0.55pt}).}
	\end{figure}

	Rewriting the model's density \eqref{eq:ModelJanossy}, for example in the marked case \eqref{eq:markedU} and \eqref{eq:markedV}, we have 
	\begin{align*}
	&j(\omega)
	=C\prod_{(x,i,m)\in\omega}\exp\Biggl[\beta_{i,0}
	+\sum_{k=1}^K\beta_{i,k}X_k(x)\Biggr]\\
	&\ \times\prod_{i_2=1}^p\prod_{z=(x_1,i_1,m_1)\in\omega}\max_{\eta\in S((\omega\setminus\{z\})_{i_2},N)}\prod_{(x_2,i_2,m_2)\in\eta}\exp\Biggl[\alpha_{i_1,i_2}\varphi_{R^{\mathrm S}_{i_1,i_2}}\biggl(\frac{2\,\|x_1-x_2\|}{m_1+m_2}\biggr)\Biggr]\\
	&\ \times\prod_{i_2=1}^p\prod_{z=(x_1,i_1,m_1)\in\omega}\max_{\eta\in S((\omega\setminus\{z\})_{i_2},N)}\prod_{(x_2,i_2,m_2)\in\eta}\exp\Biggl[\gamma_{i_1,i_2}\psi_{R^{\mathrm M}_{i_1,i_2}\leftrightarrow R^{\mathrm L}_{i_1,i_2}}\biggl(\frac{2\,\|x_1-x_2\|}{m_1+m_2}\biggr)\Biggr].
	\end{align*}
	When $N=\infty$, this is precisely a pairwise interaction Gibbs point process (see e.g., \cite[Section~6.2]{MW}) with inhomogeneous intensity for species $i$ given by 
	\begin{equation*}
	\exp\Biggl[\beta_{i,0}+\sum_{k=1}^K\beta_{i,k}X_k(x)\Biggr],\qquad x\in W,
	\end{equation*}
	and pairwise interaction functions
	\begin{equation}
	\label{eq:PairwiseInteractionFunctions}
	\exp\Biggl[2\alpha_{i_1,i_2}\varphi_{R^{\mathrm S}_{i_1,i_2}}\biggl(\frac{2\,\|x_1-x_2\|}{m_1+m_2}\biggr)+2\gamma_{i_1,i_2}\psi_{R^{\mathrm M}_{i_1,i_2}\leftrightarrow R^{\mathrm L}_{i_1,i_2}}\biggl(\frac{2\,\|x_1-x_2\|}{m_1+m_2}\biggr)\Biggr]
	\end{equation}
	(the factor $2$ in front of $\alpha_{i_1,i_2}$ and $\gamma_{i_1,i_2}$ respectively, arises because for any pair $x_1$, $x_2$ of locations in $\omega$, our model double-counts the pairwise interaction between $x_1$ and $x_2$).
	Equation \eqref{eq:PairwiseInteractionFunctions} above makes clear the joint effect of the short and medium range potentials, as well as the effect of the magnitude and sign of the coefficients $\alpha_{i_1,i_2}$ and $\gamma_{i_1,i_2}$.
	A plot to illustrate this effect is provided in Figure~\ref{fig:sum_potentials}.
	
	When $N$ is finite, the model only accounts for interactions between each individual and its $N$ closest neighbours.
	This explains our use of the adjective `saturated' to qualify our model.
	
	\subsection{Some cases of interest}
	
	\subsubsection{Non-interacting model}
	
	Assume that $\alpha_{i_1,i_2}=\gamma_{i_1,i_2}=0$, so that there is neither attraction nor repulsion.
	Our general model \eqref{eq:ModelJanossy} simplifies to
	\begin{equation*}
	j(\omega)=C\prod_{(x,i,m)\in\omega}\exp\Biggl[\beta_{i,0}
	+\sum_{k=1}^K\beta_{i,k}X_k(x)\Biggr],
	\end{equation*}
	which can be seen (see e.g., \cite{DV}) to be a multi-type inhomogeneous Poisson point process with intensity for the $i$-th type given by
	\begin{equation*}
	\exp\Biggl[\beta_{i,0}
	+\sum_{k=1}^K\beta_{i,k}X_k(x)\Biggr],\qquad x\in W.
	\end{equation*}
	In other words, each of the species is modelled independently by inhomogeneous Poisson point processes with log-intensities driven linearly by the environmental covariates.
	
	\subsubsection{Multivariate Geyer model}
	
	We assume now that $\beta_{i,k}=0$ and $\gamma_{i_1,i_2}=0$.
	We further assume that the short range interaction potential is the step potential from Table~\ref{tab:potentials}.
	The density in the unmarked case \eqref{eq:unMarkedU} is equal to
	\begin{multline*}
	j(\omega)
	=C
	\exp\Biggl[\sum_{i=1}^p|\omega_i|\beta_{i,0}
	+\sum_{i_2=1}^p\sum_{z=(x_1,i_1,m_1)\in\omega}\alpha_{i_1,i_2}\\
	\times\min\bigl(N,\bigl|\bigl\{(x_2,i_2,m_2)\in(\omega\setminus\{z\})_{i_2}\;:\;\|x_1-x_2\|\le R^{\mathrm S}_{i_1,i_2}\bigr\}\bigr|\bigr)\Biggr],
	\end{multline*}
	which is an instance of the class of models used in \cite{RMO}.
	
	\section{Inference}
	\label{sec:inference}
	
	\subsection{Logistic regression of Baddeley et al. (2014)}
	\label{subsec:logistic}
	
	In this subsection, we prove that the assumptions of \cite{BCRW} hold, which ensures that their logistic regression can be used to do inference for our model.
	This method enables us to estimate the parameters $\beta$, $\alpha$ and $\gamma$.
	
	\begin{sloppypar}
	The density of the model defined in \eqref{eq:ModelJanossy} can be written as
	\begin{equation}
	\label{eq:ExponentialFamilyJanossy}
	j(\omega)
	=C\exp\bigl(\theta^\intercal t(\omega)\bigr).
	\end{equation}
	In the equation above, we have defined the parameter vector $\theta:=(\theta_0^T,\theta_1^T,\theta_2^T,\theta_3^T)^T$, where $\theta_0:=(\beta_{1,0},\dots,\beta_{p,0})^T$, $\theta_1:=(\beta_{1,1},\dots,\beta_{1,n},\ldots,\beta_{p,1},\ldots,\beta_{p,n})^T$, $\theta_2:=(\alpha_{1,1},\dots,\alpha_{p,p})^T$ and $\theta_3:=(\gamma_{1,1},\dots,\gamma_{p,p})^T$.
	\end{sloppypar}
	
	In addition, we have set $t(\omega):=(t_0(\omega)^T,t_1(\omega)^T,t_2(\omega)^T,t_3(\omega)^T)^T$, where \begin{equation*}
	t_0(\omega):=(|\omega_1|,\dots,|\omega_p|)^T,\qquad t_1(\omega):=(s_1(\omega)^T,\dots,s_p(\omega)^T)^T,
	\end{equation*}
	\begin{equation*}
	t_2(\omega):=(s_{1,1}(\omega)^T,\dots,s_{p,p}(\omega)^T)^T,\qquad t_3(\omega):=(\widetilde s_{1,1}(\omega)^T,\dots,\widetilde s_{p,p}(\omega)^T)^T,
	\end{equation*}
	for 
	\begin{align*}
	&\qquad\qquad\qquad s_i(\omega):=\biggl(\sum_{(x,i,m)\in\omega}X_1(x),\dots,\sum_{(x,i,m)\in\omega}X_K(x)\biggr)^T,\\
	&\qquad\qquad\qquad s_{i,j}(\omega):=\sum_{z=(x,i,m)\in\omega}u(z,(\omega\setminus\{z\})_{j}),
	\end{align*}
	and
	\begin{equation*}
	\widetilde s_{i,j}(\omega):=\sum_{z=(x,i,m)\in\omega}v(z,(\omega\setminus\{z\})_{j}).
	\end{equation*}
	Under this new compact notation \eqref{eq:ExponentialFamilyJanossy}, the Papangelou conditional intensity at $\omega\in\mathcal N$ and for an individual of species $i\in\{1,\dots,p\}$ with mark $m$ located at $x\in W$ is readily computed as
	\begin{equation}
	\label{eq:ExponentialFamilyPapangelou}
	\pi((x,i,m),\omega)
	=\exp\bigl(\theta^\intercal t((x,i,m),\omega)\bigr),
	\end{equation}
	where $t((x,i,m),\omega):=t(\omega\cup\{(x,i,m)\})-t(\omega)$.
	
	The fact that we can write the density and the Papangelou conditional intensities respectively as \eqref{eq:ExponentialFamilyJanossy} and \eqref{eq:ExponentialFamilyPapangelou} guarantees that the assumptions of \cite{BCRW} hold.
	Given an observed configuration $\omega$, the logistic regression technique of \cite{BCRW} can be summarised as:
	\begin{enumerate}
		\item
		sample a set of dummy points $D$ with known (fixed) intensity, denoted by $\rho$;
		\item
		compute $t(z,\omega\setminus\{z\})$ defined in \eqref{eq:ExponentialFamilyPapangelou} as $z$ ranges over $\omega\cup D$;
		\item
		obtain $\theta$ defined above by a logistic regression with response variable $1$ when $z=(x,i,m)\in\omega$ and $0$ otherwise, input variables $t(z,\omega\setminus\{z\})$ and offset term $-\log(\rho(x))$.
	\end{enumerate}
	
	\subsection{Variance-Covariance matrix}
	\label{subsec:stderrors}
	
	Our model belongs to the class of Gibbs point processes and as such, standard errors and confidence intervals are not straightforward to produce. 
	Indeed, it has been shown in \cite{BCRW} that, although the standard errors corresponding to the logistic regression of the previous section are a good approximation, they are in general not accurate. 
	Instead, asymptotic confidence intervals can be estimated by the technique introduced in \cite{CR13} (see also Section~4 and the appendices of \cite{BCRW}).
	We will not repeat here the details of the construction of the asymptotic variance-covariance matrix, but we draw the reader's attention to the fact that there appears to be multiple typographic errors in equation~(A4) of \cite{BCRW}.
	We refer to our package described in the supplementary material for the details of the implementation.
	
	\subsection{Estimation of the other parameters}
	\label{subsec:OtherEstimation}
	
	Section~\ref{subsec:logistic} dealt with the estimation of $\beta$, $\alpha$ and $\gamma$.
	It remains to explain how to choose the saturation parameter $N$, the shape of the potential functions, as well as the interaction radii between and within species on the short, medium and long ranges.
	
	We shall often fix the potential shapes in order to simplify the analysis.
	Regarding the saturation parameter $N$, in some cases, we shall keep it fixed to $2$.
	This assumption implies that the probability of a new individual being at a given location depends only on its two neighbours with which it interacts most, disregarding other individuals.
	Another option would be to follow the last paragraph of Section~2.2 in \cite{RMO} and set $N$ automatically depending on the observed abundances.
	
	In \cite{RMO}, the interaction radii are fixed a priori, and they write as their justification ``in data analysis one usually has a priori information on relevant ranges (e.g., \cite{UCCH})''.
	Although a priori fixing these parameters has been done in some of our analyses, we also wanted a straightforward statistical procedure to estimate the interaction radii.
	This has allowed us to fit the model to different datasets without prior knowledge of the characteristics of the species involved.
	
	Our basic idea is to calibrate the model for various values of the interaction radii, saturation parameters, and potential shapes, and choose the set of values which performs best according to some measure of goodness of fit.
	Since one of our goals is to apply the model to large-scale datasets, an important requirement for the measure of goodness-of-fit is that it be relatively fast to compute.
	Consequently, we have refrained from using computationally heavy techniques like that of \cite{MB} or an explicit computation of the likelihood as in Section~8.3.2 of \cite{MW}.
	Instead, we propose as a measure of the goodness of fit the pseudo-likelihood corresponding to the logistic regression in Section~\ref{subsec:logistic}.
	More explicitly, we choose values of the saturation parameter and interaction radii which maximise the logistic pseudo-likelihood.
	
	\section{Simulation}
	\label{sec:simulation}
	
	\subsection{Coupling from the past}
	\label{subsec:CFTP}
	
	In some cases, it is possible to use the `coupling from the past' algorithm (sometimes called `perfect simulation' algorithm) to sample from our point process, see Section~11 of \cite{MW}.
	Contrary to other simulation algorithms, the `coupling from the past' algorithm is not approximate, and produces samples from the actual point process.
	In order to apply such an algorithm in practice, one needs to prove that its Papangelou conditional intensity is locally stable, i.e., that there exists a function $h$ such that $\pi((x,i,m),\omega)\le h(x)$ almost everywhere.
	The following Proposition~\ref{prop:LocallyStable} ensures that our model is locally stable under some additional hypotheses.
	We define $x^+:=\max(x,0)$ for any real number $x$.
	\begin{prop}
		\label{prop:LocallyStable}
		Assuming that for any $i_1,i_2$, $\gamma_{i_1,i_2}\le0$, we have
		\begin{equation*}
		\pi((x,i,m),\omega)\le h_1(x,i),
		\end{equation*}
		for almost any $x\in W$, $1\le i\le p$, $m\in\mathbb R$ and $\omega\in\mathcal N$, and where
		\begin{equation*}
		h_1(x,i):=\exp\Biggl[\beta_{i,0}
		+\sum_{k=1}^K\beta_{i,k}X_k(x)
		+6N\sum_{j=1}^p\alpha_{i,j}^+\Biggr].
		\end{equation*}
	\end{prop}
	\begin{proof}
		The proof is a straightforward consequence of Lemma~1 in the supplementary material.
	\end{proof}
	
	Given Proposition~\ref{prop:LocallyStable} above, we shall often work under the assumption 
	\begin{equation*}
	\text{\bfseries(H)}\quad\gamma_{i_1,i_2}\le0,\quad1\le i_1,i_2\le p,
	\end{equation*}
	which is to say that none of the medium-range interactions are attractive.
	Under {\bfseries(H)}, Proposition~\ref{prop:LocallyStable} ensures that the `coupling from the past' algorithm can be applied.
	The details of how the algorithm applies to our setting are provided in Section~3 of the supplementary material.
	
	\subsection{Metropolis-Hastings algorithm}
	\label{subsec:mh}
	
	Although the algorithm introduced in the previous subsection is extremely powerful, it has two disadvantages.
	First, it is sometimes slow, and for some values of the parameters, it does not converge in a reasonable time.
	Second, it requires the additional hypothesis {\bfseries(H)} which we would like to relax in some instances.
	As such, in some cases, we will fall back on the unconditional Metropolis-Hastings algorithm, see Algorithm~7.4 of \cite{MW}.
	There are a series of possible variations of the algorithm, see for example Remark~7.6 of \cite{MW} for a specialisation to the locally stable setting.
	
	Since we aim for a version of the algorithm which can be applied to simulate from our model in all settings, we shall choose, in the notation of \cite{MW}, a probability of birth equal to $1/2$, uniformly distributed births $q_b(\cdot)=\mathbf1_{W}(\cdot)/|W|$, and a probability $1/2$ of a uniformly distributed death distributed according to $q_d(\cdot,\omega)=\mathbf1_{\omega}(\cdot)/|\omega|$, where $\omega\in\mathcal N$.

	\section{Numerical simulations}
	\label{sec:numerical}
	
	\subsection{Simulation study}
	\label{subsec:FirstSimulation}
	
\begin{figure}[!ht]
		\centering
		\includegraphics[trim={0cm 0cm 6.5cm 0},clip,scale=0.27]{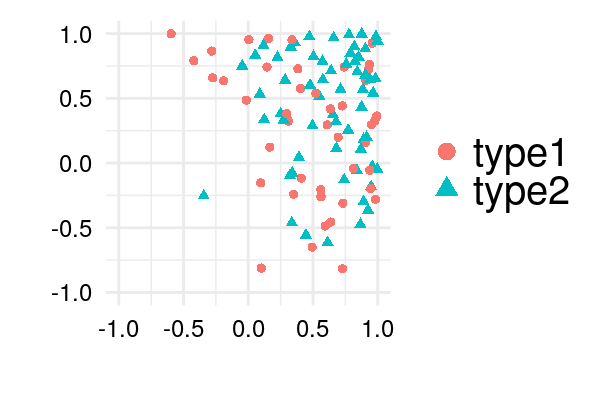}
		\includegraphics[trim={0cm 0cm 6.5cm 0},clip,scale=0.27]{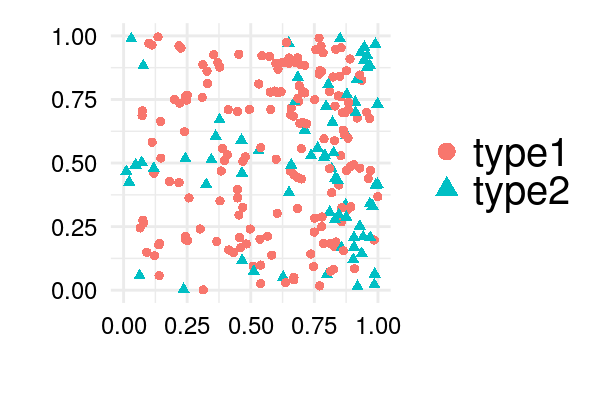}
		\includegraphics[trim={0cm 0cm 0cm 0},clip,scale=0.27]{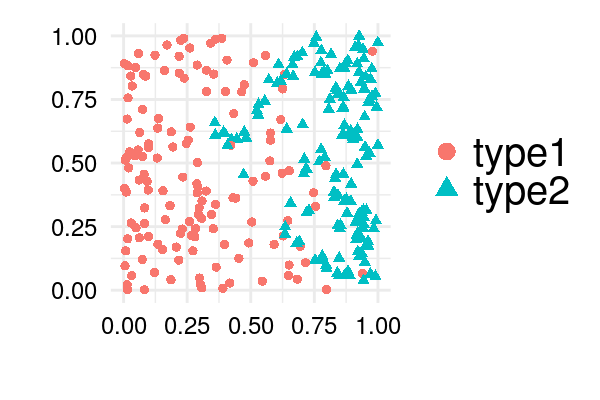}
		\caption{\label{fig:typical_sample} Typical samples considered in our numerical experiments.
		On the left, a sample from the point process considered in Section~\ref{subsec:FirstSimulation}, in the middle, a sample from the point process considered in Section~\ref{subsec:sensitivity} and finally on the right, a sample from the point process considered in Section~\ref{subsec:unknown_radius}.}
	\end{figure}

	We start with a simulation study involving two species.
	This ensures that the number of parameters is tractable, while still demonstrating that the `coupling from the past' algorithm and the fitting procedure are working as expected.
	We ran simulation studies involving significantly more species, and we have not observed any decrease in performance.
	We report the results of a seven species study in the supplementary material.
	In this first numerical experiment, we consider a `saturated pairwise interaction Gibbs point process' on the square region $W=[-1,1]^2$, consisting of $p=2$ species, with no marks, and whose distribution is driven by two geospatial covariates, $X_1(x,y)=x$ and $X_2(x,y)=y$.
	We consider uniform short-range interaction radii of $R^{\mathrm S}=0.05$, medium-range interaction radii of $R^{\mathrm M}=0.07$ and long-range interaction radii of $R^{\mathrm L}=0.12$.
	The rest of the parameters are given by $\beta_0^T=(2.5,2)$, $\beta_1^T=(2,2.5)$ (corresponding to $X_1$), $\beta_2^T=(1,1.5)$ (corresponding to $X_2$), and
	\begin{equation*}
	\alpha=\begin{pmatrix}
	-0.2 & 0.1\\
	0.1 & -0.6
	\end{pmatrix},\qquad
	\gamma=\begin{pmatrix}
	-0.6 & -0.3\\
	-0.3 & 0
	\end{pmatrix}.
	\end{equation*}
	We set the saturation parameter $N$ to 2, take as the short-range potential the square bump function, and choose the normal medium-range potential, see Table~\ref{tab:potentials}.
	In order to illustrate our experiment, we plot on the left of Figure~\ref{fig:typical_sample} a typical sample from this point process.
	
	We sampled $1,000$ independent draws of this point process.
	Since the assumption {\bfseries(H)} from Section~\ref{subsec:CFTP} is satisfied and the simulation procedure is reasonably fast for these parameters, these draws are sampled by the `coupling from the past' algorithm.
    The saturation parameter, interaction distances, and interaction shapes were set to their true values.
	We then fit each of the samples by the logistic regression technique from Section~\ref{subsec:logistic}, and produced asymptotic confidence intervals according to Section~\ref{subsec:stderrors}.
	The results are presented in Table~\ref{tab:coverage}.
	Our results are satisfying, showing good mean estimates over only $1,000$ samples, along with coverage probabilities with a mean and median of 95\%.
	
	\begin{table}
	\small 
		\caption{\label{tab:coverage} Parameter estimates \& coverage probabilities.}
		\centering
		\begin{tabular}{|M{0.14\textwidth}|M{0.12\textwidth}|M{0.11\textwidth}|M{0.12\textwidth}|M{0.12\textwidth}|M{0.18\textwidth}|}
			\hline
				Parameter & True value & Mean & Median & RMSE & Coverage prob.\\
			\hline
			$\beta_{1,0}$ & 2.50 & 2.50 & 2.51 & 0.334 & 0.94 \\
            $\beta_{2,0}$ & 2.00 & 1.97 & 1.98 & 0.341 & 0.94 \\
            $\beta_{1,1}$ & 2.00 & 2.14 & 2.10 & 0.575 & 0.95 \\
            $\beta_{2,1}$ & 2.50 & 2.67 & 2.63 & 0.567 & 0.97 \\
            $\beta_{1,2}$ & 1.00 & 1.08 & 1.04 & 0.418 & 0.96 \\
            $\beta_{2,2}$ & 1.50 & 1.62 & 1.58 & 0.448 & 0.95 \\
            $\alpha_{1,1}$ & -0.20 & -0.53 & -0.34 & 1.12 & 0.95 \\
            $\alpha_{1,2}$ & 0.10 & 0.092 & 0.11 & 0.273 & 0.95 \\
            $\alpha_{2,2}$ & -0.60 & -0.82 & -0.75 & 0.562 & 0.95 \\
            $\gamma_{1,1}$ & -0.60 & -0.69 & -0.66 & 0.397 & 0.96 \\
            $\gamma_{1,2}$ & -0.30 & -0.31 & -0.30 & 0.179 & 0.96 \\
            $\gamma_{2,2}$ & 0.00 & -0.018 & -0.023 & 0.0239 & 0.94 \\
			\hline\end{tabular}
	\end{table}
	
	\subsection{Sensitivity analysis}
	\label{subsec:sensitivity}
	
	In this experiment, we study how sensitive our calibration is to mis-specified values of the interaction radii and the saturation parameter $N$.
	We consider a `saturated pairwise interaction Gibbs point process' on $W=[0,1]^2$, consisting in $p=2$ species, with no marks, and whose distribution is driven by a single environmental covariate $X_1(x,y)=x$.
	We assume that the two species interact over different ranges, and that their distribution is characterised by $\beta_0^T=(4,3.5)$, $\beta_1^T=(1.5,2)$, and
	\begin{equation*}
	R^{\mathrm S}=\begin{pmatrix}
	0.04 & 0.06\\
	0.06 & 0.03
	\end{pmatrix},\qquad\alpha=\begin{pmatrix}
	0.4 & -0.3\\
	-0.3 & 0.4
	\end{pmatrix},\qquad\gamma=\begin{pmatrix}
	0 & 0\\
	0 & 0
	\end{pmatrix}.
	\end{equation*}
	We take as the short-range potential the square bump function from Table~\ref{tab:potentials}, and choose a saturation parameter $N=2$. 
	
	Although the assumption {\bfseries(H)} from Section~\ref{subsec:CFTP} is satisfied, it is faster to sample $100$ independent draws of this point process by the Metropolis-Hastings algorithm of Section~\ref{subsec:mh}, with $100,000$ steps.
	In order to give a sense of the type of point process we are working with, we show in the middle of Figure~\ref{fig:typical_sample} a typical sample.
	
	In our experiment, we first fit each of the samples by mis-specifying the short-range interaction radii $R^{\mathrm S}$, then assumed a mis-specification of the saturation parameter $N$.
	More specifically, we consider two mis-specifications of the interaction radii, namely
	\begin{equation*}
	R^{\mathrm S}_-=R^{\mathrm S}-\begin{pmatrix}
	0.02 & 0.02\\
	0.02 & 0.02
	\end{pmatrix}\quad\text{ and }\quad
	R^{\mathrm S}_+=R^{\mathrm S}+\begin{pmatrix}
	0.02 & 0.02\\
	0.02 & 0.02
	\end{pmatrix}.
	\end{equation*}
	We also consider an under-specified saturation parameter $N_-=1$ and an over-specified $N_+=4$.
	
	\begin{table}
		\caption{\label{tab:misspecification} Mis-specification of the interaction radii as either an under-specification $R^{\mathrm S}_-$ or an over-specification $R^{\mathrm S}_+$.}
		\centering
		\begin{tabular}{|M{0.12\textwidth}|M{0.12\textwidth}|M{0.07\textwidth}|M{0.19\textwidth}|M{0.07\textwidth}|M{0.19\textwidth}|}
			\multicolumn{1}{c}{}& \multicolumn{1}{c}{} &\multicolumn{2}{c}{$\overbracket[0.6pt][2pt]{\rule{0.28\textwidth}{0pt}}^{\text{\normalsize Under-specification}}$} & \multicolumn{2}{c}{$\overbracket[0.6pt][2pt]{\rule{0.28\textwidth}{0pt}}^{\text{\normalsize Over-specification}}$} \\
			\hline
			Parameter& True value & Mean & Coverage prob. & Mean & Coverage prob. \\
			\hline
			$\beta_{1,0}$ & 4.0 & 4.16 & 0.80 & 3.77 & 0.92 \\\hline
			$\beta_{2,0}$ & 3.5 & 3.41 & 0.90 & 3.51 & 0.97 \\\hline
			$\beta_{1,1}$ & 1.5 & 1.64 & 0.93 & 1.64 & 0.93 \\\hline
			$\beta_{2,1}$ & 2.0 & 2.35 & 0.87 & 2.06 & 0.93 \\\hline
			$\alpha_{1,1}$ & 0.4 & 0.38 & 0.92 & 0.44 & 0.97 \\\hline
			$\alpha_{1,2}$ & -0.3 & -0.38 & 0.87 & -0.31 & 0.97 \\\hline
			$\alpha_{2,2}$ & 0.4 & 0.52 & 0.90 & 0.30 & 0.93 \\\hline
		\end{tabular}
	\end{table}
	
	The results for the interaction radii mis-specification are presented in Table~\ref{tab:misspecification}.
	The main insight gained from this part of the experiment is that the estimates of the parameters are fairly accurate even when the interaction radii have been mis-specified by around 50\%.
	This is largely due to the shape of our short-range potential function which is flat around the origin, and the high intensity of points in each sample relative to the saturation parameter $N$.
	In addition, we remark that the estimates are notably better when the radius is mis-specified as $R^{\mathrm S}_+$. 
	Our interpretation of this fact is that when the user chooses an interaction radius which is larger than the true one, the same broad pairwise interactions are accounted for.
	When the radius is under-specified instead, some pairwise interaction are strongly discounted, which biases the estimates of some of the parameters.
	
	\begin{table}
		\caption{\label{tab:misspecification_saturation} Mis-specification of the saturation parameter as either $N_-$ or $N_+$.}
		\centering
		\begin{tabular}{|M{0.12\textwidth}|M{0.12\textwidth}|M{0.07\textwidth}|M{0.18\textwidth}|M{0.07\textwidth}|M{0.18\textwidth}|}
			\multicolumn{2}{c}{} &\multicolumn{2}{c}{$\overbracket[0.6pt][2pt]{\rule{0.28\textwidth}{0pt}}^{\text{\small Under-specification}}$} & \multicolumn{2}{c}{$\overbracket[0.6pt][2pt]{\rule{0.28\textwidth}{0pt}}^{\text{\small Over-specification}}$} \\
			\hline
			Parameter& True value & Mean & Coverage prob. & Mean & Coverage prob. \\
			\hline
			$\beta_{1,0}$ & 4.0 & 4.06 & 0.88 & 4.03 & 0.91 \\\hline
			$\beta_{2,0}$ & 3.5 & 3.51 & 0.94 & 3.41 & 0.91 \\\hline
			$\beta_{1,1}$ & 1.5 & 1.69 & 0.89 & 1.53 & 0.96 \\\hline
			$\beta_{2,1}$ & 2.0 & 2.19 & 0.90 & 2.10 & 0.94 \\\hline
			$\alpha_{1,1}$ & 0.4 & 0.52 & 0.91 & 0.22 & 0.48 \\\hline
			$\alpha_{1,2}$ & -0.3 & -0.54 & 0.62 & -0.19 & 0.53 \\\hline
			$\alpha_{2,2}$ & 0.4 & 0.35 & 0.96 & 0.28 & 0.90 \\
			\hline\end{tabular}
	\end{table}
	
	The results related to the mis-specification of the saturation parameter $N$ are in Table~\ref{tab:misspecification_saturation}.
	A few things stand out in this analysis.
	First, the $\beta$ parameters (which relate to the abundance) are well estimated even when the saturation parameter is mis-specified.
	Indeed, the mean estimated values of $\beta_{1,0}$, $\beta_{2,0}$, $\beta_{1,1}$, $\beta_{2,1}$, $\beta_{1,2}$ and $\beta_{2,2}$ are very close to the true values, and the associated coverage probabilities are of the right magnitude.
	Second, some interaction coefficients have very bad coverage probabilities, but broadly speaking their signs and magnitude are properly recovered by the estimation procedure. 
	Third, when the saturation parameter is under-specified, the corresponding interaction coefficients are larger in magnitude, while when it is over-specified the interaction coefficients are smaller. 
	Heuristically, this is due to the fact that when the saturation parameter is under-specified, there are less interactions accounted for in the sum of short-range interactions \eqref{eq:unMarkedU}, and consequently the corresponding interaction coefficient that multiplies the sum ought to be larger.
	
	\subsection{Inference of the interaction radii}
	\label{subsec:unknown_radius}
	
	In this paragraph, we assume that the true interaction radii are unknown, and we study how well the model is able to recover them using our proposed method from Section~\ref{subsec:OtherEstimation}.
	We do not choose the same parameters as in the previous Section~\ref{subsec:sensitivity} since, as observed there, the model is not very sensitive to the actual value of the interaction radius.
	Instead, we purposely choose strong interaction coefficient values to allow our fitting procedure to recover the true values of the interaction radii.
	
	We choose an observation window $W=[0,1]^2$, with $p=2$ species, no marks, and whose distribution is driven by a single geospatial covariate $X_1(x,y)=x-0.5$.
	We assume that all interactions occur at a distance of $0.05$ and in addition we assume that the interactions at those ranges are quite strong, so that the calibration procedure is able to pick them up.
	To be explicit, the rest of the parameters are given by $\beta_0^T=(6.5,2.6)$, $\beta_1^T=(-1,1)$, and
	\begin{equation*}
	\alpha=\begin{pmatrix}
	-1 & -0.5\\
	-0.5 & 2
	\end{pmatrix},\qquad\gamma=\begin{pmatrix}
	0 & 0\\
	0 & 0
	\end{pmatrix}.
	\end{equation*}
	We choose $N=2$ for the saturation parameter and take as the short-range potential the exponential function from Table~\ref{tab:potentials}.
	A typical sample is shown on the right of Figure~\ref{fig:typical_sample}.
	
\begin{figure}[!ht]
		\centering
		\includegraphics[trim={0cm 0cm 0cm 0cm},clip,width=0.95\textwidth]{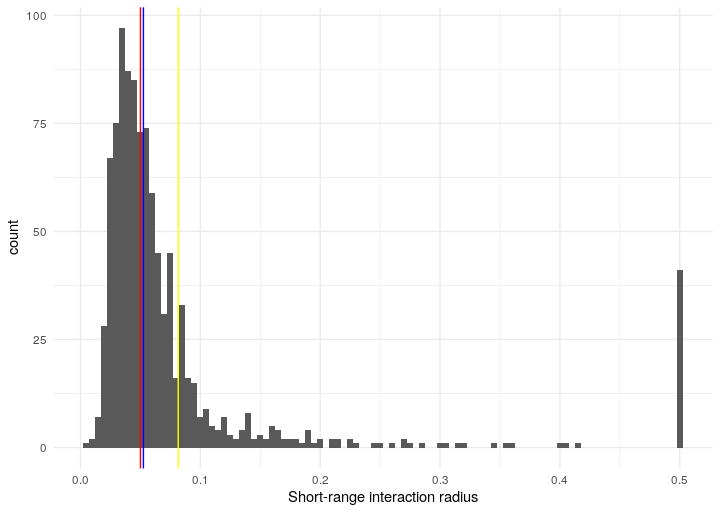}
		\caption{\label{fig:count} Optimal short-range interaction radius for each draw, obtained by pseudo-likelihood maximization.
		The maximization was done on a discrete grid between $0.0025$ and $0.5$.
		The true value of the interaction radius is shown in red, the median estimate is in blue, and the average estimate (including the values hitting the hard limit at $0.5$) is drawn in yellow.}
	\end{figure}

	We sampled $1000$ independent draws of this point process.
	Although the assumption {\bfseries(H)} from Section~\ref{subsec:CFTP} is satisfied, these draws are sampled with $1,000,000$ steps of the Metropolis-Hastings algorithm which is quicker for such extreme values of the interaction coefficients.
	For each draw of the point process, we find the optimal short-range interaction coefficient by maximising the pseudo-likelihood. 
	We find in Figure~\ref{fig:count} that for around 4\% of samples, the pseudo-likelihood is actually maximised by choosing the largest possible interaction radius.
	When removing these outliers, the mean estimated short-range interaction radius is found to be $0.06$.
	If instead we keep these samples, then the mean estimate significantly overestimates the true interaction radius, and the median actually works best.
	
\begin{figure}[!ht]
		\centering
		\includegraphics[trim={0cm 0cm 0cm 0cm},clip,width=0.95\textwidth]{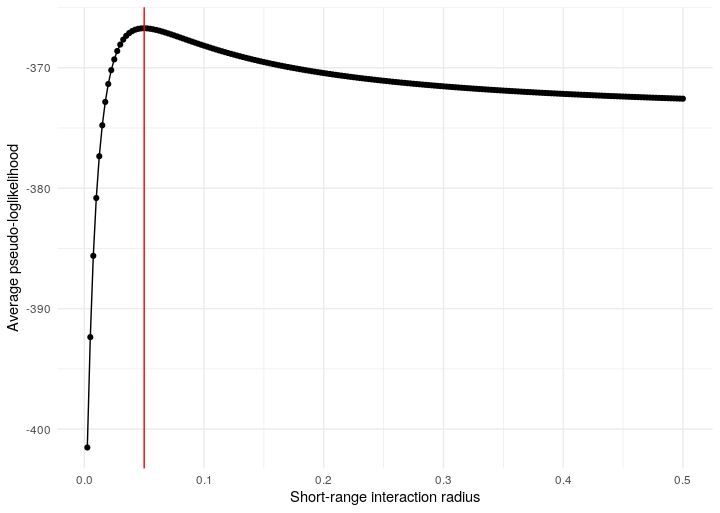}
		\caption{\label{fig:loglik} Pseudo-loglikelihood averaged over all samples, for a given value of the short-range interaction coefficient.
		The value that maximizes the average log-likelihood is found to be the true value of the interaction radius, $R^{\mathrm S}$=0.05, shown here in red.}
	\end{figure}

	In order to explore how well our method is actually performing, we also searched for the interaction radius which maximises the average pseudo-likelihood over all draws. 
	Although not practical, since this method requires the observation of multiple replications of the point process, we show in Figure~\ref{fig:loglik} that this technique properly recovers the true value of the short-range interaction radius.
	
	Although we have reported here the results of a study with quite extreme values of the interaction coefficients, our reported findings are representative of a range of other tested values.
	In running the simulation with other interaction coefficients, we find that the main change is in the proportion of samples for which the method does not properly converge.
	We found this proportion to vary between $4\%$ and $30\%$.
	We gather from this experiment that the method introduced in Section~\ref{subsec:OtherEstimation} works reasonably well to estimate unknown interaction radii, except in certain cases where the pseudo-likelihood maximising radius appears to be infinite.
	In conclusion, we caution the reader to not put much confidence in estimated values of the interaction radii hitting the hard upper-bound, especially when the corresponding interaction coefficient is not statistically significant.
	
	\section{Real applications}
	\label{sec:datasets}
	In this section, we consider three different case studies from plant ecology.
	In each case we give examples of ecological insights derived from our model.
	All three datasets consist of the locations of trees, differing however in their biome, plot size, density of individuals and number of species.
	
	\subsection{Norway spruces}
	\label{subsec:spruces}
	
\begin{figure}[!ht]
	\centering
	\includegraphics[trim={0 0 0 0},width=\textwidth]{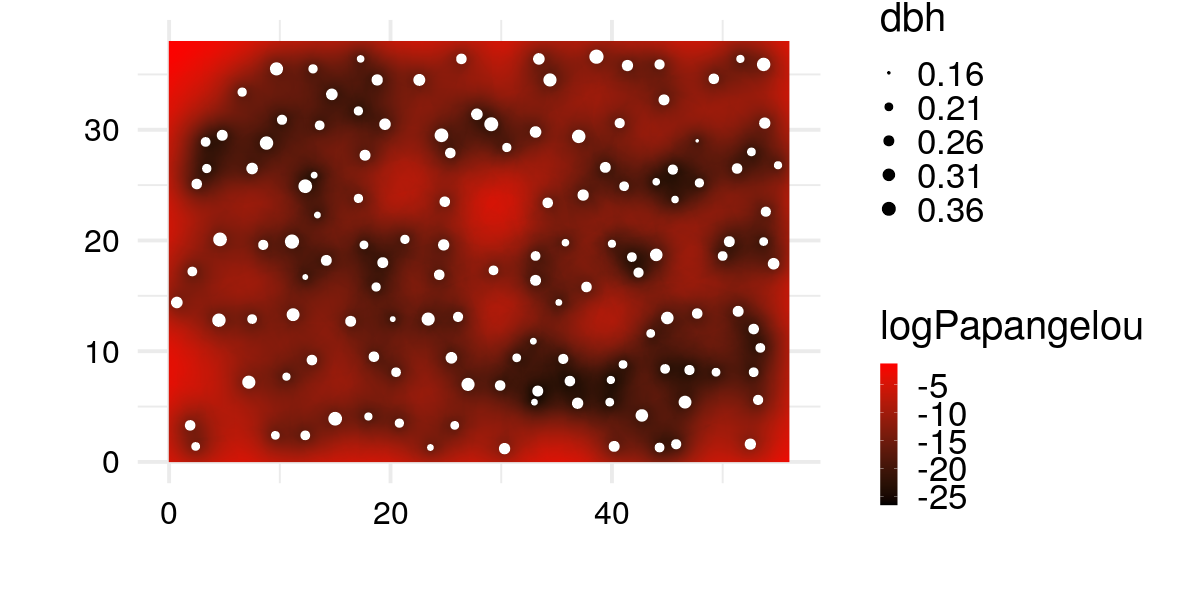}
		\caption{\label{fig:spruces} Norway spruces with marks representing their diameter at breast height.
			The background colour gradient is the fitted log-Papangelou conditional intensity.}
	\end{figure}

	In this subsection we consider the locations of 134 Norway spruce trees in a natural forest stand in Saxonia, Germany. 
	The original source of the data is unknown, but it has been widely studied in the point process literature, see for example Section~4 of \cite{F} and Example~2 in \cite{GSG}.
	The diameter at breast height in meters has been recorded for each individual tree in the dataset, and will serve as our marks.
	There are no associated environmental covariates, and instead the dataset is often used as an example of a regular marked point process, with interaction distances thought to be proportional to marks.
	What we call interaction radii are sometimes described in the literature on this dataset as ``influence zones''~\citep{GSG}, ``hard-core'' and ``interaction'' radii~\citep{PSH}.
	Various estimates of these values have been derived in previous analyses and one of our aims shall be to compare our results to the literature.
	In Figure~\ref{fig:spruces}, we show the locations of the spruces along with discs proportional to their diameters.
	
	\subsubsection*{Results}
	
	Following \cite{GSG}, we assume that interactions take place at distances proportional to the marks, and so we choose \eqref{eq:markedU} and \eqref{eq:markedV}, which in words assumes that individual to individual interactions are proportional to the average marks of the two individuals.
	In order to estimate the interaction radii, potential function shapes and saturation parameter, we deployed the multi-dimensional maximisation outlined in Section~\ref{subsec:OtherEstimation}, using the pseudo-likelihood of the logistic regression as the objective function.
	Our only constraint is restricting the saturation parameter to the range of values $\{1,2,4,6\}$; however we found that the fit was not significantly influenced by these values.
	The results of our model are summarised in Table~\ref{tab:spruces}.
	
	\begin{table}
		\caption{\label{tab:spruces} Norway spruce dataset results.
			We do not give the 95\% confidence intervals for the parameters fitted by the ad-hoc pseudolikelihood maximisation.
			The other confidence intervals are produced by the method outlined in Section~\ref{subsec:stderrors}.}
		\centering
		\begin{tabular}{|c|c|c|c|}
			\cline{2-4}
			\multicolumn{1}{c|}{}&Parameter& Estimate & 95\% CI\\
			\hline
			Intercept & $\beta_0$ & $-1.88$ & $(-2.57, -1.19)$ \\	\hline	
			Short-range coefficient & $\alpha$ & $-5.18$ & $(-6.92, -3.43)$ \\\hline
			Medium-range coefficient & $\gamma$ & $0.14$ & $(0.05, 0.23)$ \\\hline
			Short-range radius & $R^{\mathrm S}$ & 2.41 & \strike{c|}{} \\\hline
			Medium-range radius & $R^{\mathrm M}$ & 16.40 & \strike{c|}{} \\\hline
			Long-range radius & $R^{\mathrm L}$ & 24.43 & \strike{c|}{} \\\hline
			Short-range shape & $\varphi_{R^{\mathrm S}}$ & Exponential & \strike{c|}{} \\\hline
			Medium-range shape & $\psi_{R^{\mathrm M}\leftrightarrow R^{\mathrm L}}$ & Geyer & \strike{c|}{} \\\hline
			Saturation & $N$ & 6 & \strike{c|}{} \\
			\hline\end{tabular}
	\end{table}
	
	Recall that the radii in Table~\ref{tab:spruces} are given as a proportion of the marks, so that for example two individuals of size $0.2\,\mathrm m$ interact on the short-range at a distance of $0.2\,R^{\mathrm S}=0.482\,\mathrm m$.
	Our fitted estimates are broadly in line with what other researchers have estimated or a prior fixed in the relevant literature, see \cite{F}, \cite{PSH} and \cite{GSG}.
	Indeed, as others have observed, there are strong negative short-range interactions between the locations of the spruces.
	In addition, the authors of \cite{PSH} choose a ``hard-core radius'' of $1\,\mathrm m$, where our short-range interaction radius amounts to $0.6\,\mathrm m$ on average (calculated as $R^{\mathrm S}$ times the average tree diameter of $25\,\mathrm{cm}$).
	We find medium-range interactions that occur at an average distance of $5.1\,\mathrm m$ (calculated as the mean of $R^{\mathrm M}$ and $R^{\mathrm L}$ times the average tree diameter), which is analogous to the quantity \cite{PSH} call an ``interaction radius'' and set to $3.5\,\mathrm m$.
	The authors in \cite{GSG} choose an influence zone of five times the diameter, which again is comparable to our fitted short-range interaction radius.
	The best short-range potential function is found to be the exponential, which is notably the shape chosen for interactions in the pairwise Gibbs point process used in \cite{PSH}.
	
	We have also gone further than some of the existing models.
	To the best of our knowledge, other models do not capture the statistically significant medium-range positive interactions in the dataset, occurring between $16$ and $24$ times the diameter at breast height.
	This property of the point pattern might be caused by a mixture of pollination and seed dispersal.
	These ecological mechanisms would tend to increase the likelihood of finding individuals surrounded by others at these medium ranges. 
	
	\subsection{South Carolina Savannah river site}
	
\begin{figure}[!ht]
		\includegraphics[trim={0 0 0 0},clip,width=\textwidth]{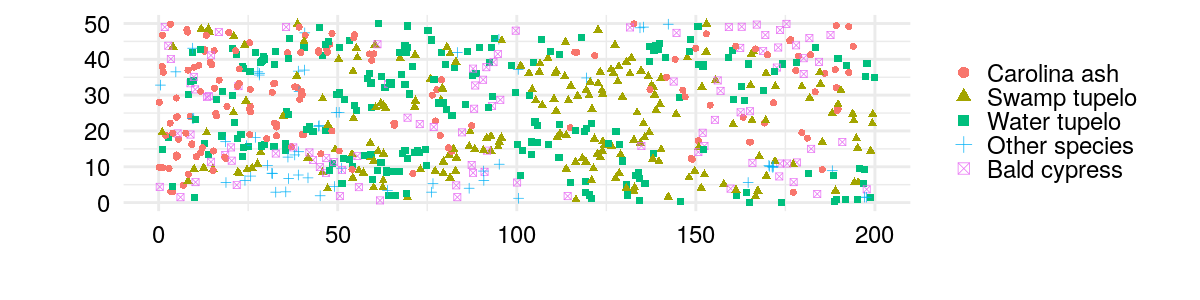}
		\caption{\label{fig:swamp} South Carolina Savannah river site.}
	\end{figure}

	In this subsection, we study the locations of 734 individual trees in a ${200\,\mathrm m\times50\,\mathrm m}$ plot in the Savannah river site, South Carolina, USA.
	Seven different plots were originally set up by Bill Good, and a first analysis of their spatial patterns was conducted in \cite{GW}, see also the subsequent analyses in \cite{JSJD} and \cite{D}.
	We focus on one of the plots from the original experiment shown in Figure~\ref{fig:swamp}.
	The dataset can be obtained using the R language (R Core Team, 2019) as \verb|ecespa::swamp| from the \verb|ecespa| package available on CRAN.
	
	There are no known environmental covariates related to this dataset, however the (unmeasured) water level is thought to be an important driver of the spatial distribution.
	Contrary to Section~\ref{subsec:spruces} and to simplify the analysis, we assume that the saturation parameter $N$ is equal to $2$, that the short-range interaction potential is the square exponential from Table~\ref{tab:potentials}, and finally we assume that there are no medium-range interactions.
	We also let the interaction radii be on a discrete grid, with grid size $1\,\mathrm m$, and constrain them to be less than $20\,\mathrm m$.

	\subsubsection*{Fitting of the interaction radii}

	In order to estimate the different interaction radii, we follow the procedure outlined in Section~\ref{subsec:OtherEstimation} and implemented in Section~\ref{subsec:spruces}.
	We find that the fitted short range interaction distances $R^{\mathrm S}$ in meters are given by
	\begin{equation*}
	\begin{matrix*}[r]
	\text{Carolina ash:}\\
	\text{Swamp tupelo:}\\
	\text{Water tupelo:}\\
	\text{Other species:}\\
	\text{Bald cypress:}
	\end{matrix*}
	\quad
	\begin{pmatrix}
	\mathbf{1} & {\transparent{0.3} 20} & {\transparent{0.3} 1} & \mathbf{5} & {\transparent{0.3} 20}\\
	{\transparent{0.3} 20} & \mathbf{3} & \mathbf{1} & \mathbf{10} & \mathbf{9}\\
	{\transparent{0.3} 1} & \mathbf{1} & \mathbf{5} & {\transparent{0.3} 20} & \mathbf{6}\\
	\mathbf{5} & \mathbf{10} & {\transparent{0.3} 20} & \mathbf{1}& {\transparent{0.3} 20}\\
	{\transparent{0.3} 20} & \mathbf{9} & \mathbf{6} & {\transparent{0.3} 20} & {\transparent{0.3} 1}
	\end{pmatrix}
	,
	\end{equation*}
	where entry $i,j$ of the matrix above corresponds to $R_{i,j}^{\mathrm S}$, the short range interaction distance between species $i$ and $j$.
	We have put in bold values of the interaction distances which are later found to be associated with significant interactions, and greyed out values which are found not to be.
    Since their corresponding interactions are weak, greyed out values carry weak statistical weight.
	In addition, values of the interaction radius attaining our hard upper-bound of $20\,\mathrm m$ should not be taken at face value given our findings in Section~\ref{subsec:unknown_radius}.
	
	We observe that the short-range interaction radii $R_{i,i}^{\mathrm S}$ within each of the species has a mean of around $2\,\mathrm m$ while the interaction radii $R_{i,j}^{\mathrm S}$ between species are on average five times larger.
	Thus, the intra-species and inter-species short-range interaction radii appear to relate to different underlying ecological processes.
	The intra-species interaction radii $R_{i,i}^{\mathrm S}$ might be related to the seed dispersal distance and the range within which individuals (of the same species) compete for resources. 
	The inter-species interaction radii $R_{i,j}^{\mathrm S}$ could be due to unmeasured environmental variation and/or be the range within which individuals (of different species) compete for resources.
	
	\subsubsection*{Results}
	
\begin{figure}[!ht]
		\includegraphics[trim={0 0 0 2cm},clip,height=0.17\textheight]{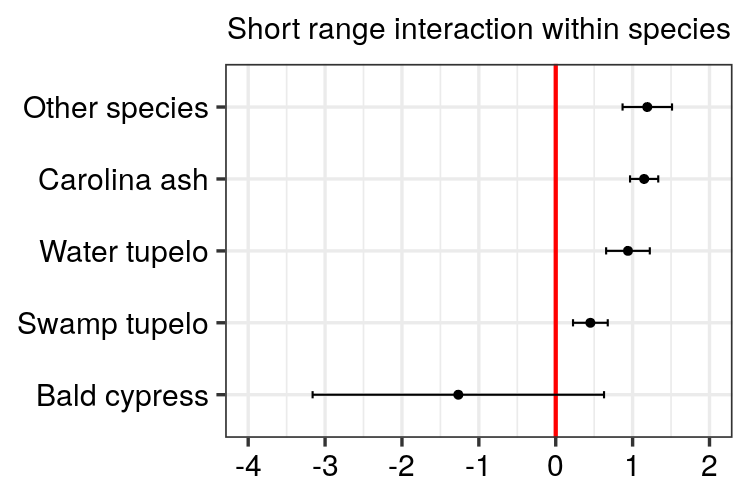}
		\includegraphics[trim={0 0 0 2cm},clip,height=0.17\textheight]{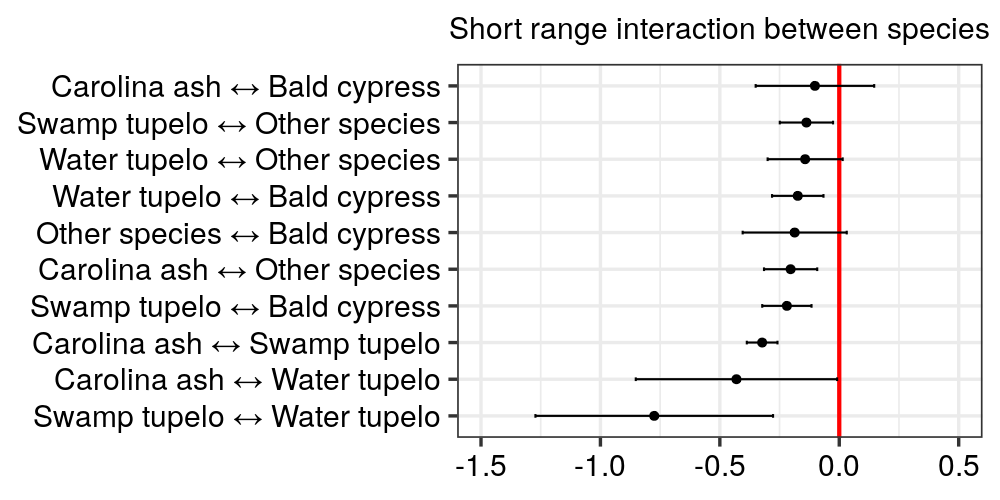}
		\caption{\label{fig:alpha} On the left-hand side, short range interaction coefficients within each of the species $\alpha_{i,i}$.
		On the right-hand side, short range interaction coefficients between each of the species $\alpha_{i,j}$.
		We provide estimates along with the corresponding $95\%$ confidence intervals.}
	\end{figure}

	The fitted values for the matrix of short-range interaction coefficients $\alpha$ are presented in Figure~\ref{fig:alpha}.
	The results support the hypothesis of strong clustering within each species, with the exception of the bald cypress in which we observe mild repulsion, although the parameter estimate is not statistically significant.
	Similar results were already obtained in \cite{D}, where it was written that the particular status of the cypress ``may be due to logging \textellipsis or it may represent some other difference between cypress and the other tree species''.
	
	The estimates of the pairwise short-range interaction radii are all negative and all but two of the 95\% confidence intervals do not overlap with zero.
	However, we recall that we have used a two-step procedure in which the interaction radii were specifically chosen to maximise the pseudo-likelihood, and in addition we have not made any correction for the multiple testing problem.
	Hence, we should be cautious in interpreting the confidence intervals.
	Broadly speaking however, there is evidence of competition rather than facilitation between species.
	We note in particular that many of the strongest repulsive associations involve the swamp tupelo. 
	These results also corroborate what was observed in the existing literature on this dataset, see in particular \cite{D}.
	However, the technique introduced in \cite{D} did not find most of the inter-species interactions to be statistically significant, perhaps due to the fact that heterogeneity in the interaction radii could not be accounted for.
	
	\subsection{Barro Colorado Island}
	\label{subsec:BCI}
	
	Fully mapped out forest plots are a rare occurrence in ecology.
	These are however crucial in understanding the relative importance of dispersal limitation, biotic interactions and habitat filtering in explaining species' distributions.
	Many seminal studies of spatial distributions within forest plots have been unable to account for inter-species associations (\cite{Condit2000, John2007PNAS_soil, wiegand2007species, shen2013quantifying}) and when they have it is via an analysis of pair correlation functions (\cite{uriarte2004spatially, Yin}).
	By contrast, our model allows us to conduct the analysis within a fully integrated model-based framework.
	
	In this section, we study the $1000\,\mathrm m\times500\,\mathrm m$ tropical moist forest plot at Barro Colorado Island, Panama.
	All woody trees and shrubs whose stems have a diameter of at least $1\,\mathrm{cm}$ have been censused in multiple years (see \cite{C}, \cite{CAMLHF} and \cite{HCF} for more details).
	Regarding the analysis of the Barro Colorado Island dataset specifically, attempts at analysing ecological drivers of multi-species distributions within a unified framework have been scarce, and we shall mostly compare our results to \cite[Section~5]{RMO} and \cite[Section~6.2]{WGJM} which are the most extensive studies to date.
	
	A wide range of environmental covariates are available for the Barro Colorado Island dataset, for example information about the soil type, elevation, etc.
	We settled upon six ecologically relevant covariates, namely slope and elevation, solar irradiance, soil pH and phosphorus content, and finally the soil moisture in the mid dry season in a non-drought year from \cite{KWER}.
	\cite{RMO} chose instead six covariates from principal component analysis, which can be difficult to interpret, while \cite{WGJM} settled on eleven different covariates including the first five of ours.
	We remark that our method scales well with the number of environmental covariates, and the reason for restricting our attention to only six of them is simply ease of presentation.
	
	There are around 300 different species and hundreds of thousands of individual trees in the Barro Colorado Island dataset, and consequently various techniques have been used to reduce the numerical complexity.
	The authors in \cite{WGJM} restrict their attention to nine seemingly arbitrarily chosen species with intermediate abundance.
	In \cite{RMO} instead, the authors exclude species for which they do not have an estimate of `reproducible size', which is used as a proxy for the size at which individuals reach reproductive maturity.
	Then for each species, the authors remove individual trees below the reproducible size threshold, and finally exclude species with less than fifty remaining individuals.
	
	In order to restrict our analysis to that of adult trees which are thought to have a more regular distribution, following \cite{RMO} we remove immature individuals from the dataset.
	Immature individuals were removed based on their size, with estimates of size at reproductive maturity available as a supplement to \cite{FOM}. 
	While \cite{RMO} exclude from their analysis the species for which the size at reproductive maturity is not available, we do not since excluding entire species from the dataset might lead to missed ecological interactions. 
	Instead, we find that reproductive maturity is well explained by a regression $Y\sim aS^b$, where $S$ is the maximum diameter of the species and $Y$ is the size at reproductive maturity.
	This leads us to exclude individuals that are below the reproductive size for their species, or if that trait is not available, below the extrapolated size at reproductive maturity inferred from their maximum diameter.
	Compared to \cite{RMO}, this retains more species.
	Finally, we group species with less than seventy individuals into a separate category which shall still play a role in the interactions accounted for by the model.
	After this procedure, we end up with 82 different species comprising around 45 thousand individual trees.
	This constitutes a few thousand more individuals and nine times more species than \cite{WGJM}; 50\% more individuals and roughly the same number of species as \cite{RMO}.
	
	We fix the saturation parameter $N$ to $2$ and let the shape of the potential functions be the square bump and normal, respectively.
	We choose $10\,\mathrm m$ as the short-range interaction radius and search for residual medium-range interactions between $20\,\mathrm m$ and $40\,\mathrm m$.
	These values are in line with the results of neighbourhood dependent growth models, see Table~4 in \cite{UCCH}.
	We implemented a Lasso regularisation of the logistic regression of Section~\ref{subsec:logistic} in order to facilitate the analysis of the many potential resulting interactions.
	The theoretical justification for using regularisation on the composite likelihood is provided in \cite{DHU}, see also \cite{BC} for the asymptotic properties of the regularised estimator in our setting.
	We chose as the regularisation parameter the one that minimises AIC. 
	
	\subsubsection*{Results}
	
	\begin{figure}[!ht]
		\includegraphics[trim={0 0 0 0},width=0.49\textwidth]{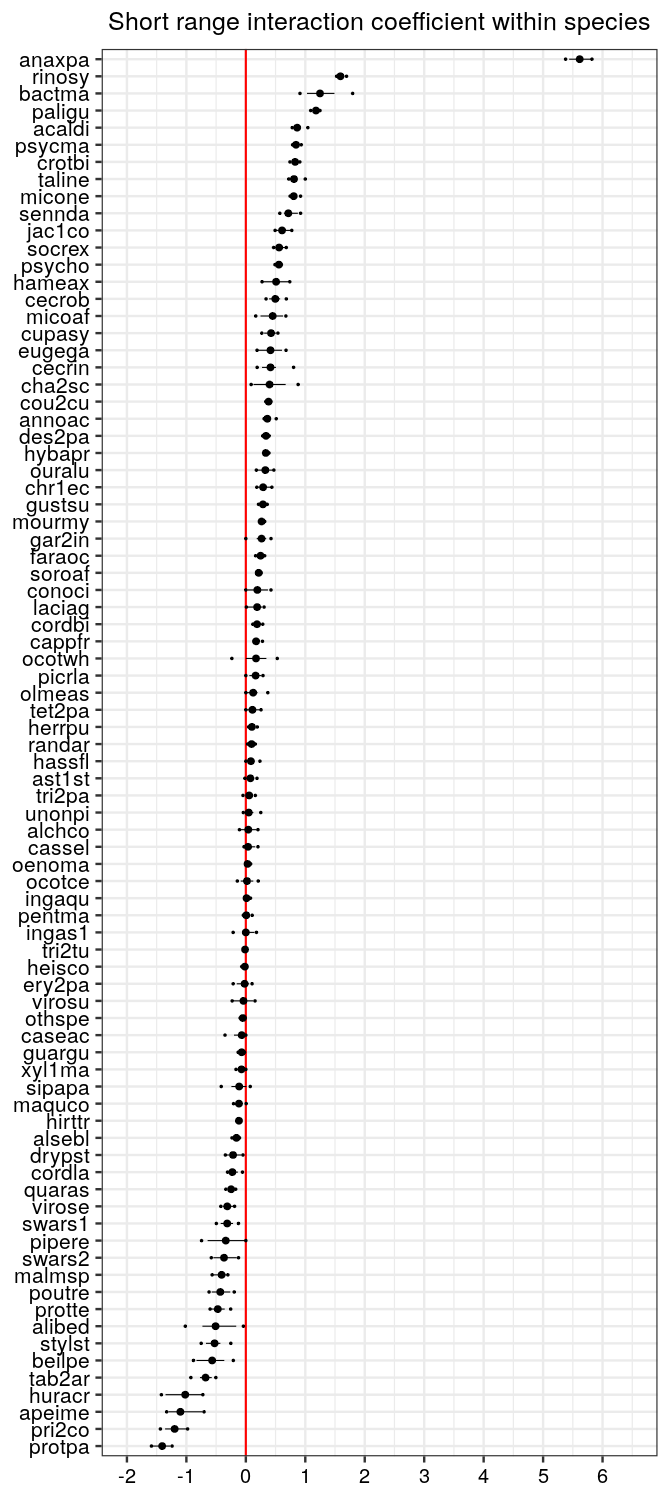}
		\includegraphics[trim={0 0 0 0},width=0.49\textwidth]{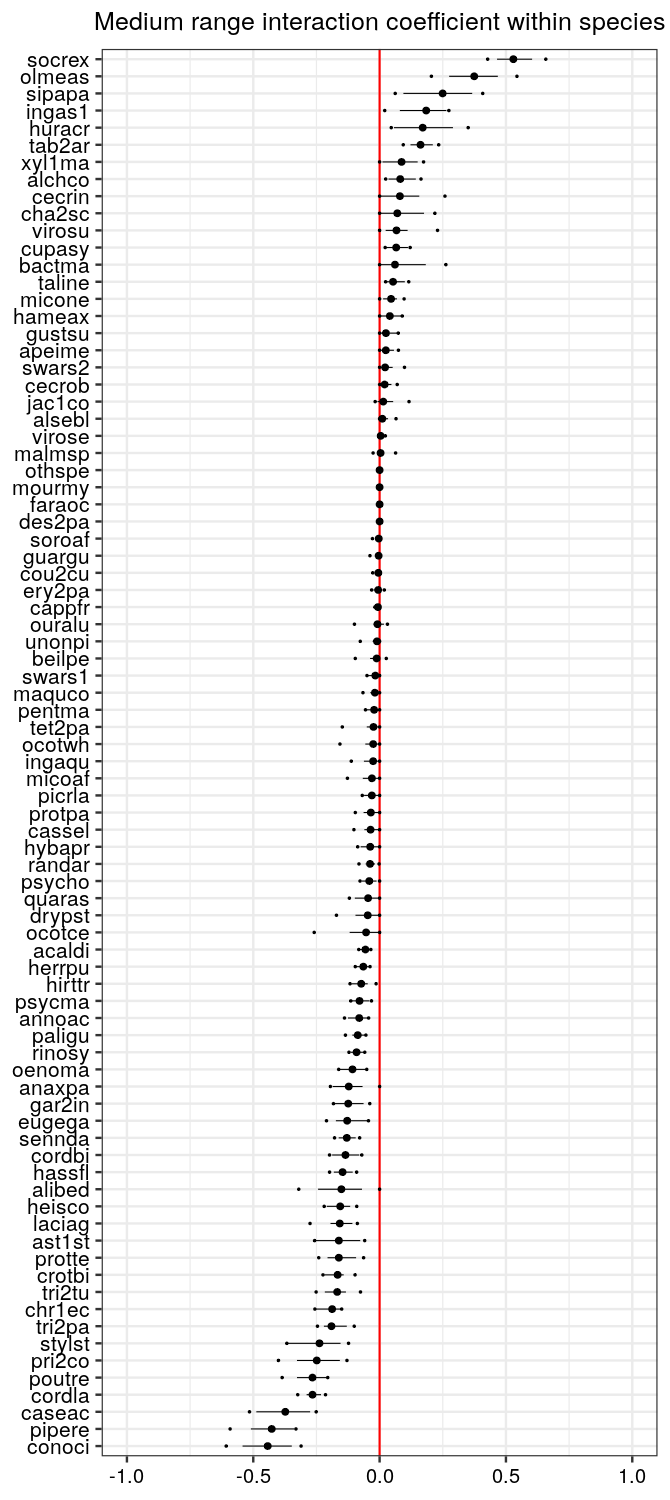}
		\caption{\label{fig:bci_within} On the left-hand side, short range interaction coefficients within each of the species $\alpha_{i,i}$.
		On the right-hand side, medium range interaction coefficients between each of the species $\gamma_{i,i}$.
		The estimates were obtained by averaging out the results of ten logistic regressions, each with a different binomial draw of the dummy points $D$.
		The error bars represent the variation among these draws.}
	\end{figure}
	
	We start by presenting in Figure~\ref{fig:bci_within} the intra-species interactions coefficients.
	We broadly observe that most species are clustered, with a few exhibiting very significant clumping.
	Notably, our three most clustered species are \textit{Anaxagorea panamensis}, \textit{Bactris major} and \textit{Rinorea sylvatica} which were highlighted in \cite{seri} as ``exceptional species'' in terms of their clustering.
	In addition, in part due to the removal of immature trees, we find some species which have negative or null intra-species interactions, leading to regular distributions.
	In Figure~\ref{fig:papangelou_4species} below we show in more detail the spatial distribution of four such species. 
	\textit{Protium panamense} is an instructive example that exhibits strong intra-species short-range negative interactions and almost no medium-range interactions.
	This species was analysed in \cite{WGJM} without removing immature trees.
	Analysing the configuration of mature trees in their framework would be more challenging since the Cox process in their model is restricted to positive associations between individuals and therefore cannot properly account for these negative intra-species interactions.
	
	\begin{figure}[!ht]
		\includegraphics[trim={0 0 0 0},clip,width=0.49\textwidth]{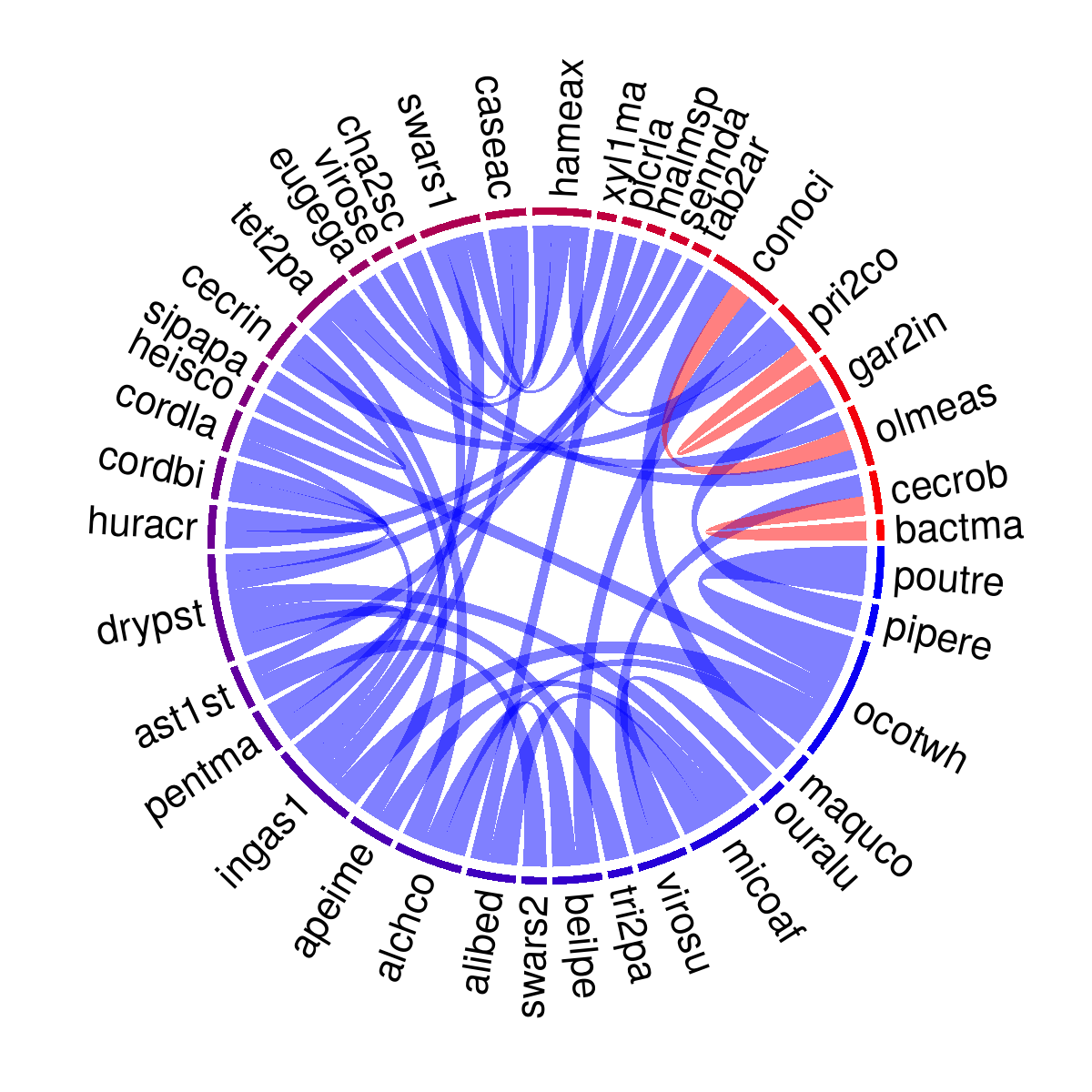}
		\includegraphics[trim={0 0 0 0},clip,width=0.49\textwidth]{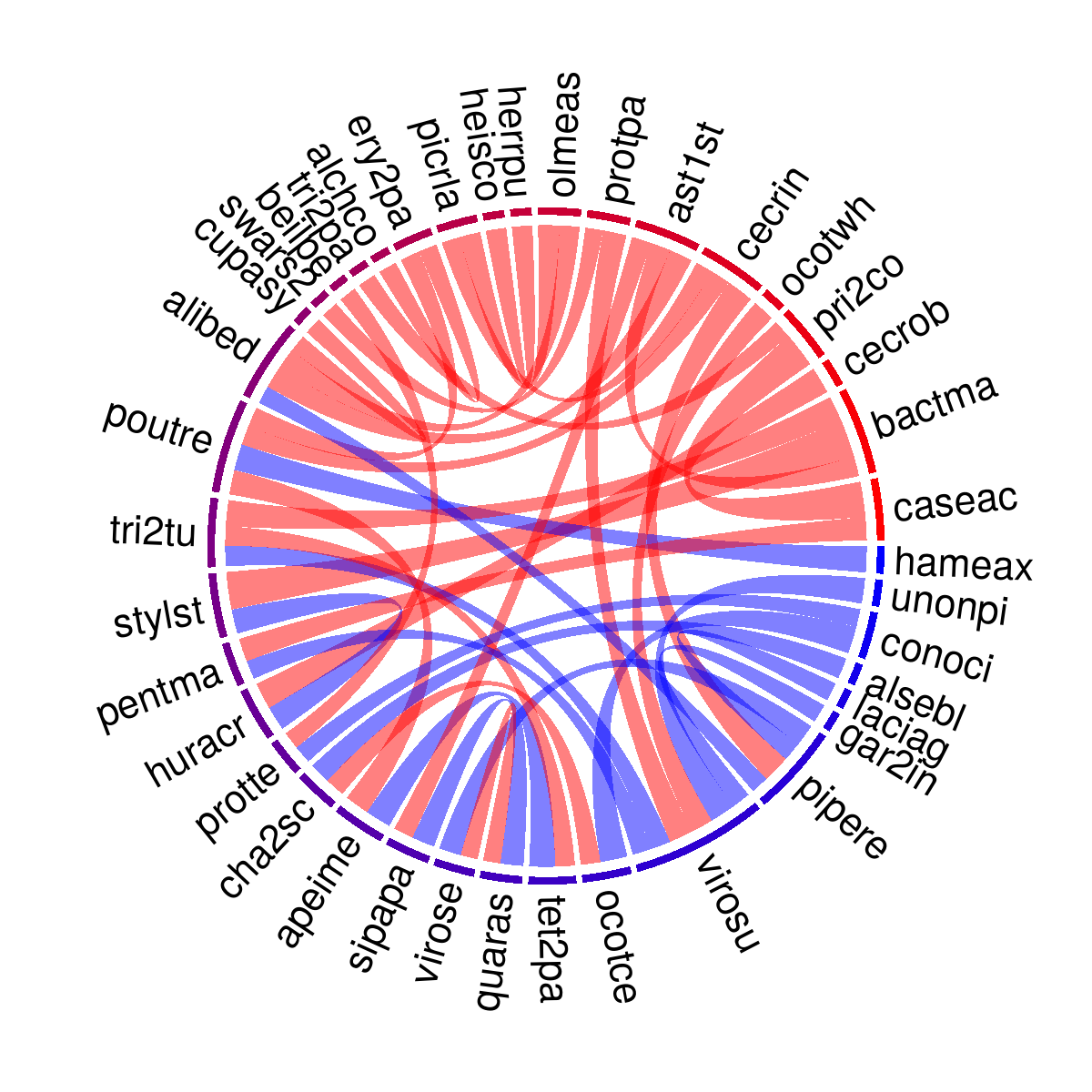}
		\caption{\label{fig:bci_between} On the left-hand side, the 40 largest short range interaction coefficients between the species $\alpha_{i,j}$.
		On the right-hand side, the 40 largest medium range interaction coefficients between each of the species $\gamma_{i,j}$.
		The coefficients shown in blue are negative, so that the corresponding interactions are repulsive, while those in red are positive, meaning the interactions are attractive.
		In both panels, the thickness of the cord is proportional to the strength of the interaction.}
	\end{figure}
	
	In Figure~\ref{fig:bci_between} we show the inter-species interaction coefficients.
	We find that our model has properly disentangled two different kinds of associations. 
	First, on the short range, species are generally negatively associated with one another, which is a strong marker of competition for resources.
	Second, on the medium range, we see substantially more positive associations, possibly indicating some dependency on unmeasured environmental covariates.
	Others in the literature (\cite{WGJM, RMO}) have not been able to disentangle these numerous short-range negative interactions from associations at broader scales.
	We find that some of the species pairs studied in \cite{WGJM} are negatively associated, for example \textit{Swartzia simplex} with most other species, or \textit{Hirtella triandra} with \textit{Garcinia intermedia}.
	These negative associations were not picked up by \cite{WGJM} while they were corroborated by our analysis of Ripley's cross $K$-function (not shown here).
	Indeed, all significant interactions in \cite{WGJM} were found to be positive.
	We were unable to compare our results with those of \cite{RMO} more closely since they did not report the species' label in their figures.
	
	Ecological processes such as dispersal and competition are expected to display distinct spatial signatures (\cite{seabloom}). 
 	We hypothesise that the outputs of the model presented here partly result from these ecological processes.
 	Our model has disentangled associations on different scales, providing a basis for dissecting the underlying ecological processes.
	
	\begin{figure}[!ht]
	    \begin{center}
		\includegraphics[width=0.85\textwidth]{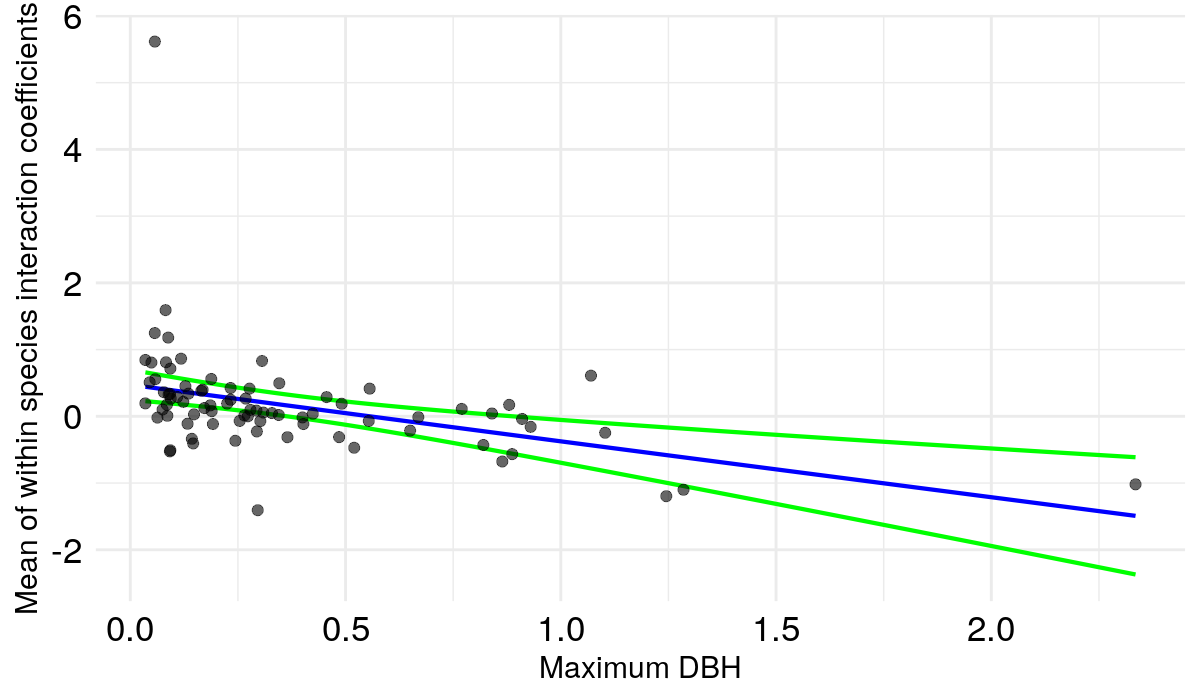}
		\end{center}
		\caption{\label{fig:bci_interaction_dbh} Mean of the intra-species interaction coefficient for each species (obtained as the average of $\alpha_{i,i}$ and $\gamma_{i,i}$) as a function of the species' maximum diameter at breast height.
		The fit shown on the figure is a GAM fit with basis dimension $3$, along with its $95\%$ confidence bands.}
	\end{figure}
	
	In terms of ecological insights, in Figure~\ref{fig:bci_interaction_dbh} we show that species with a smaller maximum diameter at breast height tend to be more clustered, with the relationship being statistically significant ($p=0.000214$ significance according to a Wald test).
	This is a well-known feature of the Barro Colorado Island dataset that our model has successfully picked up, see for example~\cite{Condit2000}.
	We also found that larger species on average have more negative associations with other species, reflecting size-dependent competitive pressure ($p<2\cdot10^{-16}$, Wald test, plot not shown here).
	
	\subsubsection*{Model assessment}
	We shall show next that our model satisfies the following compelling criteria:
	\begin{itemize}
	    \item[\textit{(i)}] for a given species, conditioning on other species and accounting for the corresponding interactions yields a conditional occurrence probability estimate which captures the inhomogeneity in the point pattern well;
	    \item[\textit{(ii)}] the intra-species interaction coefficients indicate clustering or regularity in each of the species' spatial distribution;
	    \item[\textit{(iii)}] the inter-species interaction coefficients depicted in Figure~\ref{fig:bci_between} capture actual associations between species in the dataset.
	\end{itemize}
	
	\subsubsection*{\textit{(i)} Species-specific intensity}
	
	We begin by showing that our model correctly captures the underlying spatial inhomogeneity.
	Consider a species $i$, and the configuration $\omega_{-i}$ in which we remove all individuals of species $i$. 
	Recall that the Papangelou conditional intensity $\pi((x,i,m),\omega_{-i})$ is interpreted as the probability of finding an individual of species $i$ around $x$ and with mark around $m$, conditional on individuals of other species.
	We expect individuals of species $i$ to be found at locations where this Papangelou conditional intensity takes large values. 
	We would like to assess how well the Papangelou conditional intensity $\pi((x,i,m),\omega_{-i})$ is able to separate the region into high and low density of individuals of species $i$.
	For that purpose, we compute the Area Under the ROC Curve (AUC), cf. \citet{auc}.
	In the point process framework, the AUC is computed by discretising the study area, and choosing as events the presence or absence of an individual in a cell (see, e.g., \cite{lombardo}).
	More precisely, in our context, for a (conditional) intensity $\lambda$ the AUC is defined (see Section~6.7.3 in \cite{spatstat}) as
	\begin{equation*}
	    \mathbb P(\lambda(U)<\lambda(X)),
	\end{equation*}
	where $X$ is a uniformly chosen point of the point process (in our case, of species $i$) and $U$ is a continuous random variable uniformly distributed over the study region.
	The AUC measures the ability of the intensity to properly separate the region into areas of high and low density of individuals, with a value of $0.5$ indicating a lack of discriminatory power.
	In our analysis, we have discretised the study region into $1\,\mathrm m\times1\,\mathrm m$ cells, computed $\lambda$ at each cell and at the location of each individual of species $i$ to produce an estimate of the AUC.
	We have in practice used the \verb|auc.ppp| function in \verb|spatstat|~\cite{spatstat}.
	
	\begin{figure}[!ht]
	    \begin{center}
		\includegraphics[width=0.6\textwidth]{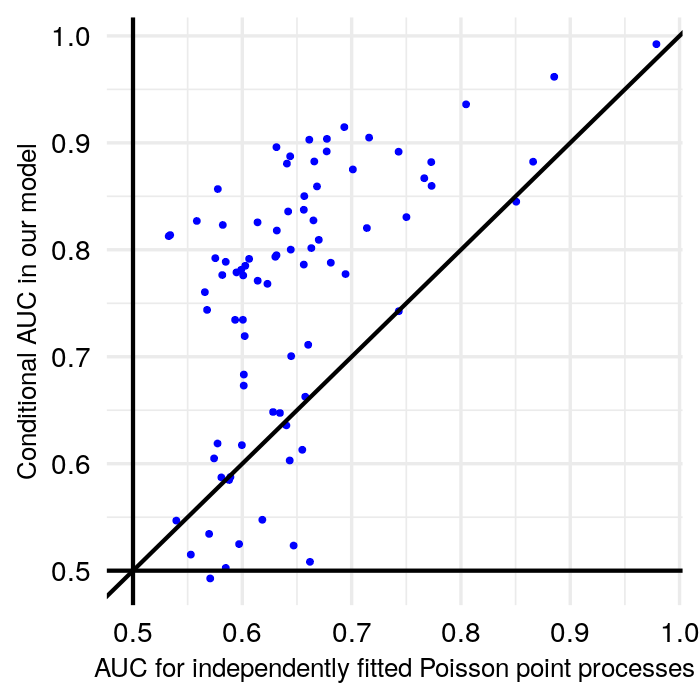}
		\end{center}
		\caption{\label{fig:conditional_auc} Conditional AUC improvement species by species, when going from an inhomogeneous Poisson point process to the saturated pairwise interaction Gibbs point process.
		Each blue point corresponds to one species.
		Points in the top-left quadrant indicate species for which our model produces a better AUC than that of an inhomogeneous Poisson point process model.
		The average AUC improvement is 0.11 and our model has improved the conditional AUC for 83\% of species.}
	\end{figure}
	
	More precisely, we proceed as follows.
    First, we fit each species separately according to a Poisson point process driven by the same six environmental covariates used in our case study, and produce a maximum likelihood intensity estimate.
    Second, for each species, we compute the Papangelou conditional intensity of our fitted Gibbs point process, conditional on other species (as described in the previous paragraph), over the whole area.
	We then compute the AUC in both cases.
	We show in Figure~\ref{fig:conditional_auc} the resulting performance gain in terms of AUC species by species.
    The saturated pairwise interactions Gibbs point process attains an average AUC of $0.76$ by conditioning on other species, compared to an average of $0.65$ for the standard Poisson point process.
    We find that the AUC of most species is improved.
    This shows that inter-species interactions are important in shaping the species' conditional distributions. 
    We acknowledge that part of this improvement is due to our model having more parameters; our main point here is that the model is indeed capturing associations between species and capitalising on these to improve the conditional intensity estimates.
    
    \begin{figure}[!ht]
    \begin{center}
        \includegraphics[width=0.65\textwidth]{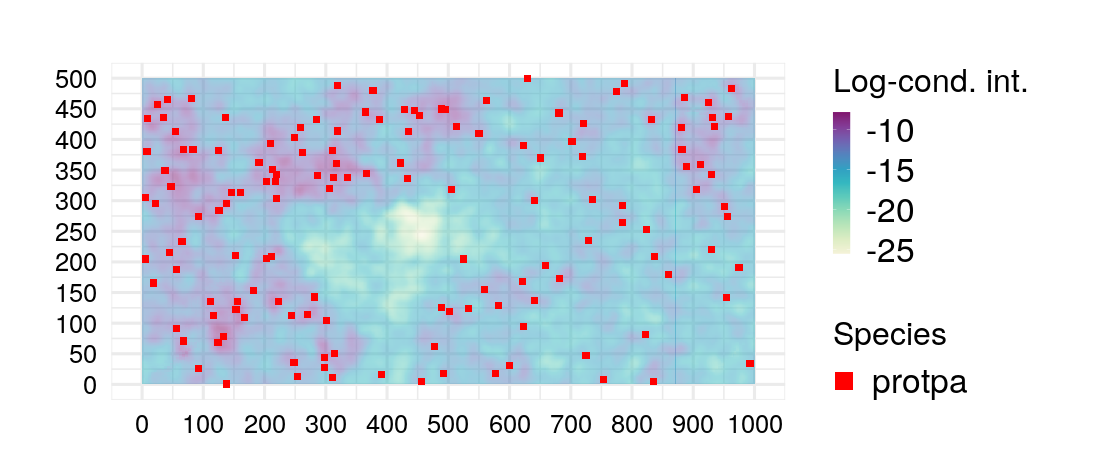}
		\includegraphics[width=0.65\textwidth]{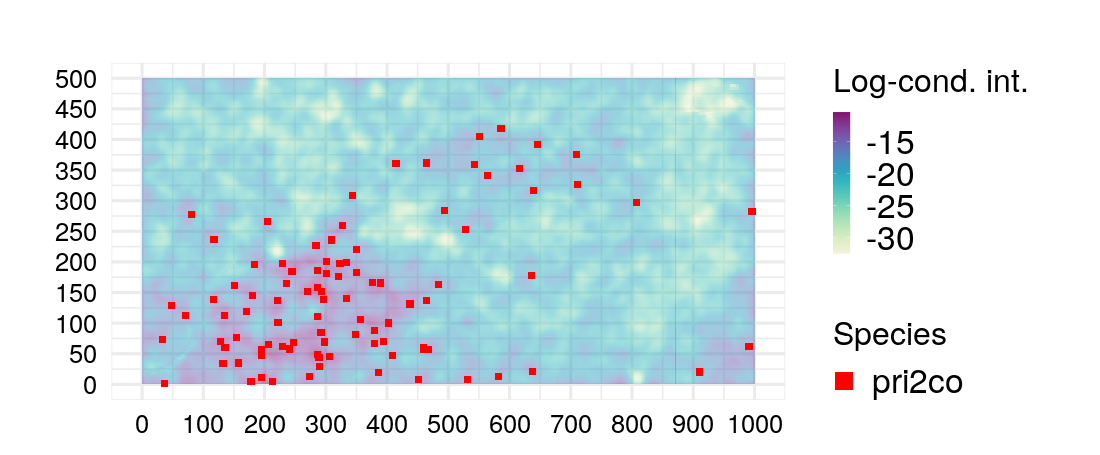}
		\includegraphics[width=0.65\textwidth]{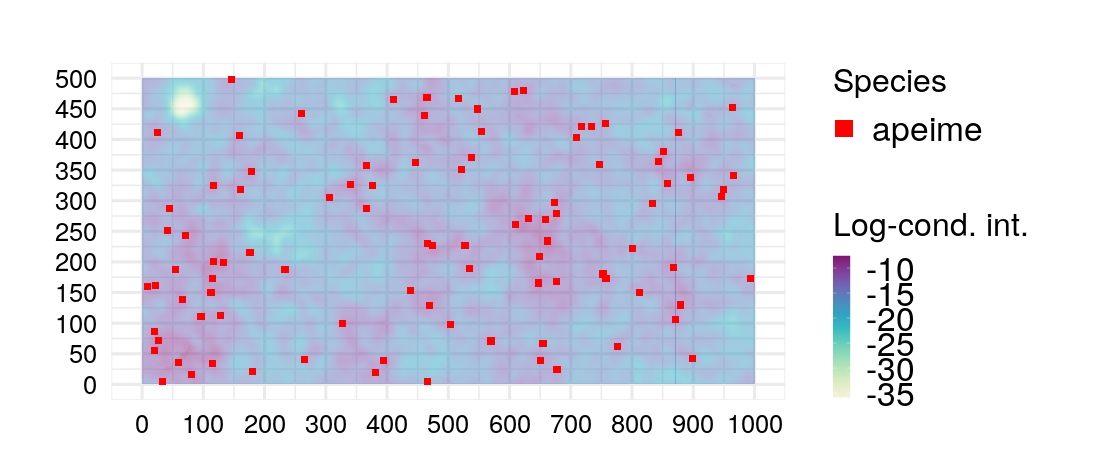}
		\includegraphics[width=0.65\textwidth]{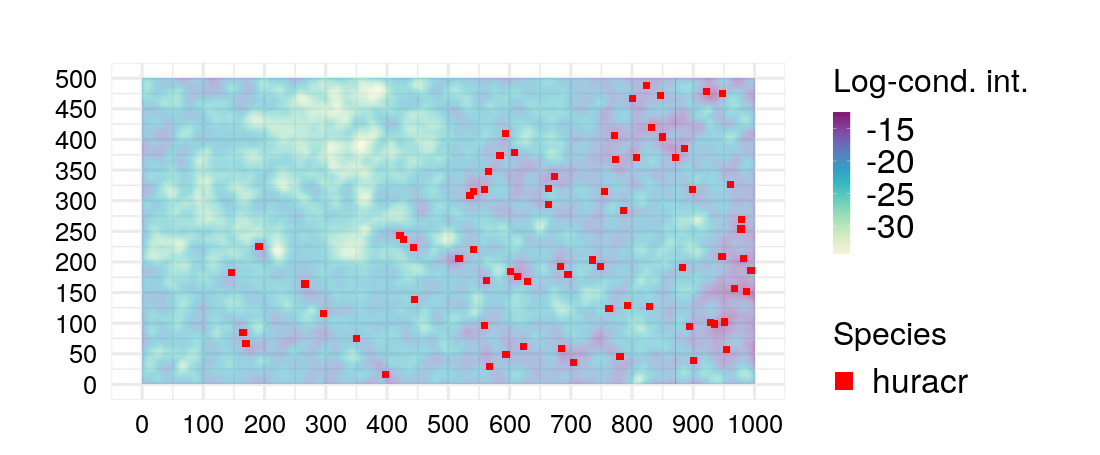}
    \end{center}
		\caption{\label{fig:papangelou_4species} Log-Papangelou conditional intensities of the four most repulsive species in our model, conditional on all other species, see the text for details on how this quantity is defined.
		Our model has captured most of the spatial inhomogeneity and its conditional intensity has properly separated the area into areas of high and low density of individuals.
		This is well quantified by the AUC metric which is quite high for these species ($\mathrm{AUC}_{\text{protpa}}=0.76$, $\mathrm{AUC}_{\text{pri2co}}=0.90$, 
		$\mathrm{AUC}_{\text{apeime}}=0.81$,
		$\mathrm{AUC}_{\text{huracr}}=0.90$).}
	\end{figure}
	
	In order to illustrate how well the Papangelou conditional intensity resembles the actual spatial distribution, let us take a closer look at the four species which were found to exhibit most intra-species short-range repulsion, namely \textit{Protium panamense} (`protpa'), \textit{Prioria copaifera} (`pri2co'), \textit{Apeiba membranacea} (`apeime') and \textit{Hura crepitans} (`huracr').
	We show in Figure~\ref{fig:papangelou_4species} the Papangelou conditional intensity computed at each of the species, conditional on other species.
	We see clearly that for these species, our model has properly separated the region into locations where the species occurs and others where it does not.
	The rather large corresponding AUC values for these species ranging from 0.76 to 0.90 corroborate this result.
	
	\subsubsection*{\textit{(ii)} Intra-species clustering}
	
	We now show that the intra-species clustering or regularity is partly captured by the intra-species interaction coefficients.
	We characterise intra-species clustering in terms of the inhomogeneous $L$-function defined, e.g., on p.~32 of \cite{MW}. 
	In general, for any two species $i$ and $j$ and a distance $R>0$, we define
	\begin{equation}
	\label{eq:AverageLFunction}
	    \overline{L_{i,j}}:=\biggl[\frac{1}{R}\int_{0}^{R}\bigl(L_{i,j}(r)-r\bigr)\,\mathrm dr\biggr]\bigg/\biggl[\frac{1}{R}\int_{0}^{R}r\,\mathrm dr\biggr]
	    =\frac{2}{R^2}\int_{0}^{R}L_{i,j}(r)\,\mathrm dr-1
	\end{equation}
	as a measure of the association between individuals of species $i$ and $j$ within a distance $R$ of each other.
	In the equation above, $L_{i,j}(r)$ is the \textit{cross} inhomogeneous $L$-function defined on p.~49 of \cite{MW}, and which generalises the usual inhomogeneous $L$-function. 

    In order to evaluate the degree of clustering in each species, we shall perform a hypothesis test (see Chapter~10 in \cite{spatstat}).
    Let our null hypothesis be that a given species $i$ is an inhomogeneous Poisson point process, conditionally on all other species. 
	By Proposition~1 in the supplementary material, this is for example the case if $\alpha_{i,i}=\gamma_{i,i}=0$ and the saturation parameter $N$ is sufficiently large.
	Under these hypotheses, the conditional point process is second order intensity reweighted stationary (SOIRS), see \cite[Definition~4.5]{MW}, and so the standard definition of $L_{i,i}$ makes sense.
	In particular, in this case $\overline{L_{i,i}}$ is expected to be zero.
    Again by Proposition~1 in the supplementary material, the intensity of the conditional point process is proportional to $\pi((x,i,m),\xi_{-i})$, where $\xi_{-i}$ is the point process consisting in individuals of species other than $i$.
    The statistic $L_{i,i}$ could be estimated by normalising the standard estimator by the fitted Papangelou conditional intensity, but we choose instead to rely on the leave-one-out kernel smoother derived in Section~2.2 of \cite{Baddeley2000}.
    If the corresponding empirical $L$-function is outside the simulation envelopes obtained by draws of an inhomogeneous Poisson point process with intensity the standard leave-one-out kernel estimate of the species, then we have grounds to reject the null hypothesis.
    When the null hypothesis does not hold, strictly speaking, the previous definition of $L_{i,i}$ and of its estimator do not make sense because first, $\pi((x,i,m),\xi_{-i})$ can not be viewed as proportional to the intensity and second, even if this were the case, the SOIRS assumption is not met. 
    However, we can expect that under the alternative the estimator of $L_{i,i}$ diverges from the expected value under the null with the same interpretation as under the SOIRS assumption.
    More precisely, values above $r\mapsto\pi r^2$ of $L_{i,i}$, and thus positive values of $\overline{L_{i,i}}$, indicate more species-specific clustering than if the species were conditionally an inhomogeneous Poisson point process.
    Negative values of $\overline{L_{i,i}}$ instead indicate more regularity.
	
	\begin{figure}[!ht]
	    \begin{center}
		\includegraphics[width=\textwidth]{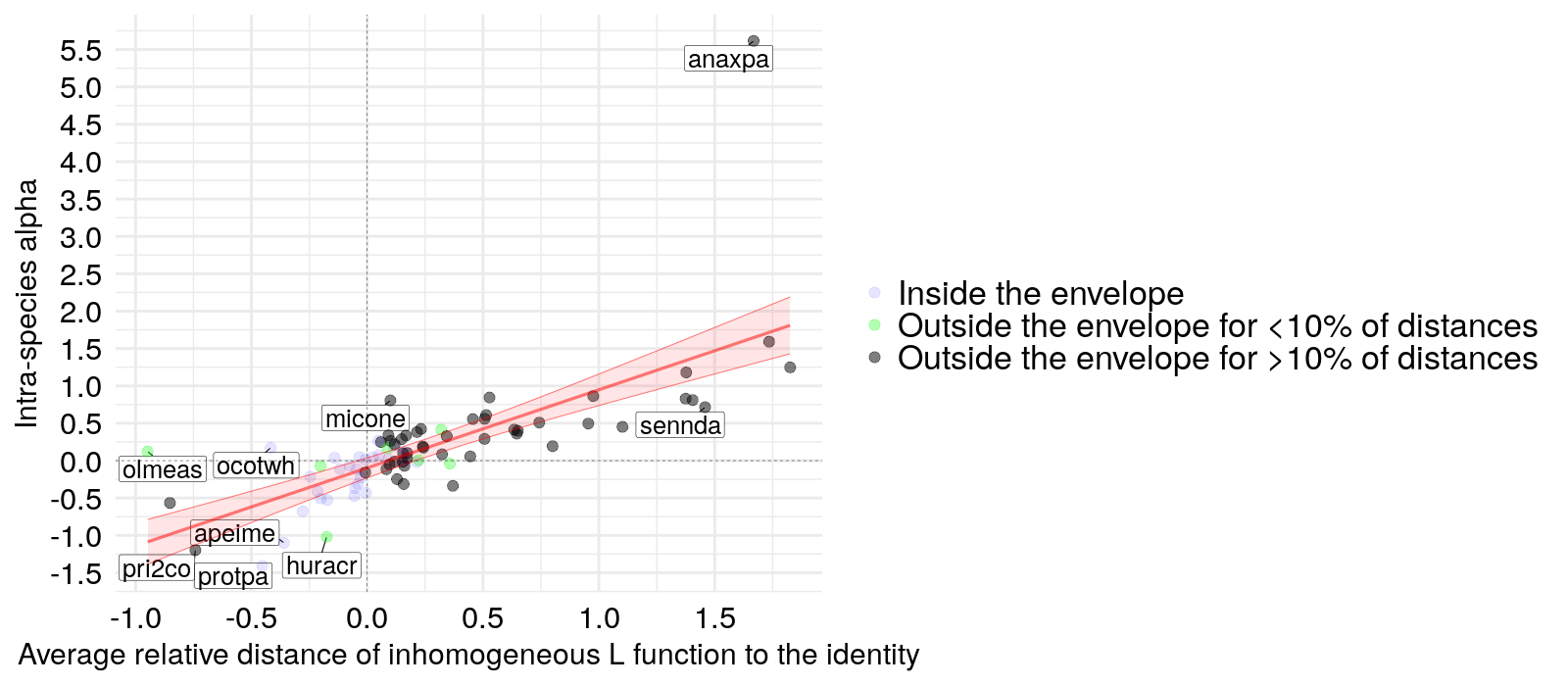}
		\end{center}
		\caption{\label{fig:within_alpha_vs_L} Scatter plot of the intra-species short-range interaction coefficients $\alpha_{i,i}$ in terms of $\overline{L_{i,i}}$, for $r$ ranging from $0\,\mathrm m$ to $20\,\mathrm m$.
		We have superimposed the results of a linear regression along with its $95\%$ confidence bands (slope $1.04$).
		The envelopes were computed with $400$ draws of an inhomogeneous Poisson point process.}
	\end{figure}
	
	We find in Figure~\ref{fig:within_alpha_vs_L} that $86\%$ of species which were above the envelopes (i.e., indicating that the species is significantly more clustered than would be expected if it were conditionally Poisson distributed) were also found to have positive short-range interaction coefficients.
	Both species which were below the envelopes were also found to have negative short-range interaction coefficients.
	In addition, we find that the intra-species short-range interaction coefficients $\alpha_{i,i}$ are positively correlated with $\overline{L_{i,i}}$, with Pearson coefficient $0.71$, and show in Figure~\ref{fig:within_alpha_vs_L} a scatter plot of all $82$ species.
	Overall, species which are more clustered than would be expected if they were conditionally Poisson distributed tend to have positive short-range intra-species interaction coefficients, and conversely species which are more regular tend to have negative coefficients. 
	This can also be seen visually in Figure~\ref{fig:papangelou_4species}, where we show that the four most repulsive species--with their estimated intensity shown in the background--tend to have a more regular distribution than that of a (conditional) inhomogeneous Poisson point process.
    
    \subsubsection*{\textit{(iii)} Inter-species clustering}
    
	We characterise inter-species associations in terms of the inhomogeneous cross $L$-function $L_{i,j}(r)$ described above.
	We still use definition \eqref{eq:AverageLFunction} to analyse inter-species interactions.
	Assume as the null hypothesis that for two species $i$ and $j$ we have $\alpha_{i,j}=\gamma_{i,j}=0$.
	By Proposition~2 in the supplementary material, the two species are independent conditionally on other species.
	By Proposition~4.4 in \cite{MW}, under these hypotheses, the conditional point process formed of the two species is cross SOIRS (see Definition 4.8 in \cite{MW}).
	In this case, the definition of $L_{i,j}$ makes sense and $\overline{L_{i,j}}$ is equal to zero.
	As in \textit{(ii)} above, strictly speaking, under the alternative hypothesis the definition of $L_{i,j}$ and its estimator do not make sense.
	However, we can again expect that values of $L_{i,j}(r)$ above their expectation under the null point to species positively associated, and conversely values below their expectation under the null indicate negatively associated species.
	Therefore, negative values of $\overline{L_{i,j}}$ correspond to repulsion and positive values correspond to positive associations (at least for small values of $r$, see p.~49 of \cite{MW}), and so this quantity serves as a good indicator of spatial associations between species.
	
\begin{figure}[!ht]
	    \begin{center}
		\includegraphics[width=\textwidth]{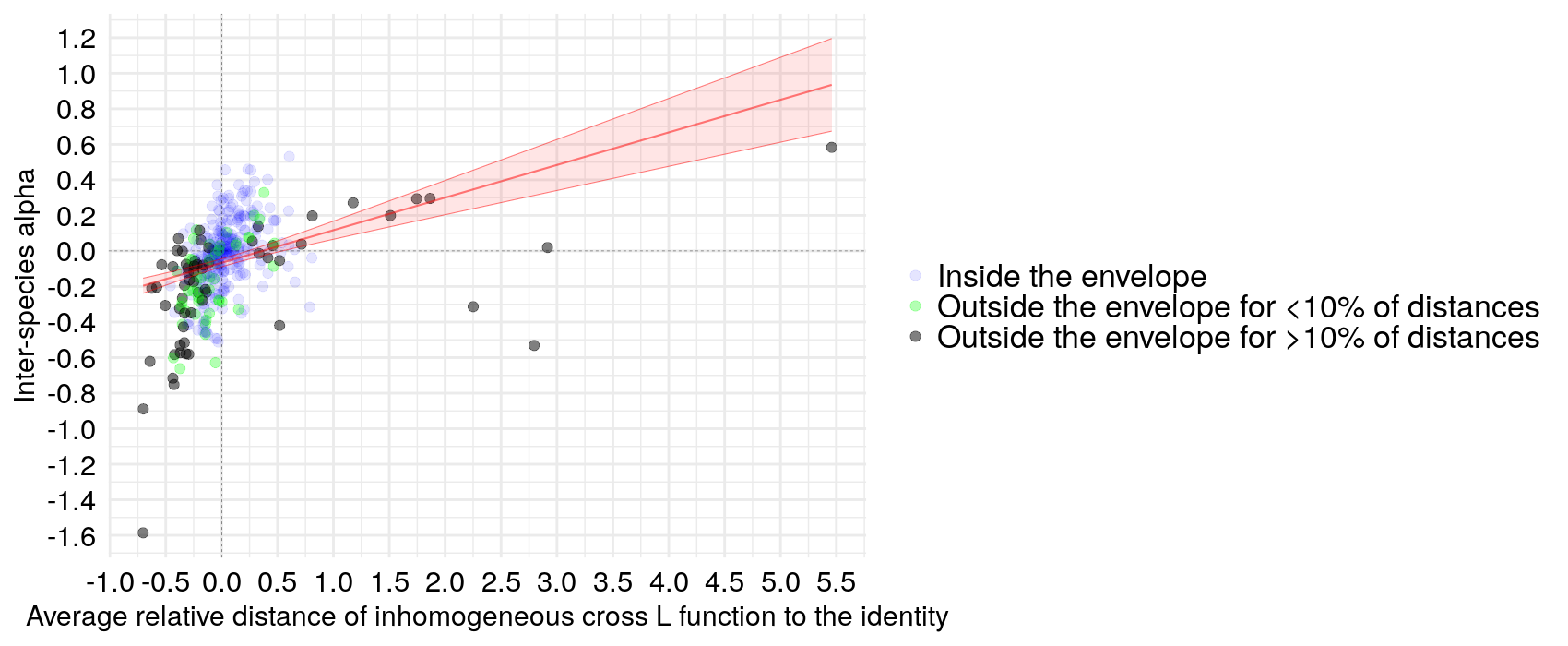}
		\end{center}
		\caption{\label{fig:between_alpha_vs_L} Scatter plot of the inter-species short-range interaction coefficients $\alpha_{i,j}$ in terms of $\overline{L_{i,j}}$, for $r$ ranging from $0\,\mathrm m$ to $20\,\mathrm m$.
		We have superimposed the results of a linear regression along with its $95\%$ confidence bands (slope $0.34$).
		The envelopes were computed with $400$ draws of two independent inhomogeneous Poisson point processes.}
	\end{figure}

	Heuristically, then, $\overline{L_{i,j}}$ represents the average relative distance to the theoretical cross $L$-function if the two species were independent conditionally on other species.
	So, for example $\overline{L_{i,j}}=-0.5$ indicates that the cross $L$-function is on average $50\%$ less than if the two species were independent.
	Envelopes are not as straightforward to produce as in the intra-species setting \textit{(ii)} above, though.
	Indeed, the null hypothesis in this case is that species $i$ and $j$ are independent point processes, but they need not be Poissonian.
	And indeed, in general they are not even saturated pairwise interaction Gibbs point process, and their simulation (conditional on other species containing tens of thousands of individuals) is very computationally demanding.
	Therefore, in Figure~\ref{fig:between_alpha_vs_L} we restrict ourselves to the 28 species which were not found in \textit{(ii)} to depart from the conditional inhomogeneous Poisson hypothesis.
	
	We find in Figure~\ref{fig:between_alpha_vs_L} that $67\%$ of species which were above the envelopes (indicating that the two species are found closer than would be expected if they were independent) were also found to have positive short-range inter-species interaction coefficients.
	In addition, $93\%$ of species which were below the envelopes were found to have negative short-range inter-species interaction coefficients.
	We also show in Figure~\ref{fig:between_alpha_vs_L} a scatter plot of all species pairs and also observe that $\overline{L_{i,j}}$ and $\alpha_{i,j}$ are positively correlated with Pearson coefficient $0.48$.
	Our findings lend credence to the fact that the short-range interaction coefficients $\alpha_{i,j}$ capture associations between individuals of different species.
	Overall, we have shown that the short-range interaction coefficients capture associations between individuals, both within and between species, and the way the model accounts for these associations convincingly models the species' conditional spatial distribution. 
	
	\section{Discussion}
	
	Two main classes of models had previously been proposed to analyse the spatial arrangement of individuals in large multi-species ecological datasets.
	First, the log-Gaussian Cox process proposed in \citet{WGJM} is an elegant model that fits within a Bayesian framework well, but cannot model competition causing repulsion within a species, nor does it scale well with the number of species.
	In addition, the latent correlated Gaussian fields have no straightforward interpretation in ecological applications.
	Furthermore, as pointed out when analysing \textit{Protium panamense} in Section~\ref{subsec:BCI}, the multivariate log-Gaussian Cox process cannot serve as a model for a species with null or negative intra-species interactions that interacts with other species.
	Second, the saturated Gibbs point process introduced in \citet{RMO} captures pairwise interactions over different ranges, and scales well with the number of species.
	We find the second class to be more compelling. 
	Inspired by the work of \citet{RMO}, in this manuscript we have introduced the `saturated pairwise interaction Gibbs point process' to start working towards a unified framework to untangle the three main drivers underlying community assembly, namely species' dispersal abilities, environmental tolerance and biotic interactions.
	
	In contrast to the model in \citet{RMO}, in modelling pairwise interactions, we allow the use of more realistic smooth potential functions instead of linear combinations of step functions.
	Moreover, our model has a role for marks such as the individuals' size, and these are thought to be influential in affecting species' distribution.
	These two features have allowed us to handle applications that are out of reach of existing models.
	For example, the locations of Norway spruces studied in Section~\ref{subsec:spruces} exhibit exponential pairwise interactions at a distance that is proportional to individuals' diameters.
	We have also studied other spatial patterns from plant ecology in which competing ecological factors are at play, and have shown how these mechanisms materialise within the framework of the model.
	We have found that our model has performed well in the Barro Colorado Island analysis in Section~\ref{subsec:BCI}, a dataset containing almost a hundred species and many thousands of individual trees.
	This has helped us gain additional insights into three very different ecosystems, namely a spruce forest from northern Europe, a subtropical swamp forest, and a neotropical rainforest.
	
	Additionally, we have addressed the problem of simulating this point process, and in particular, we proved in Proposition~\ref{prop:LocallyStable} a crucial result that allows us to apply the `coupling from the past' algorithm to draw samples from the point process.
	In our manuscript, simulating from the model has helped us carefully validate the model's performance and allowed us to do a sensitivity analysis, see Section~\ref{sec:numerical}.
	We also believe that simulating from the model will be important in future work, since it is necessary to do Monte-Carlo simulations as well as compute simulation envelopes and run goodness of fit tests.

    Our model can be applied in a wide range of settings, and may also be useful outside of ecology.
	Indeed, the notion of a physical pairwise interaction making it more or less likely that two individuals occur close by is a compelling assumption that surely also makes sense in physics, epidemiology and economics among others.
	We have consequently made our fitting and simulation procedures available as an open-source R package, see the supplementary material for more details. 

	\section*{Acknowledgements}
	
	We thank the anonymous referees whose comments helped improve a previous version of this manuscript.
	One of the referees in particular has helped us make substantial improvements to an earlier version.
	This work was supported by Australian Research Council Grant No DP190100613.

	\bibliography{library}

\begin{thebibliography}{46}
\expandafter\ifx\csname natexlab\endcsname\relax\def\natexlab#1{#1}\fi
\expandafter\ifx\csname url\endcsname\relax
  \def\url#1{\texttt{#1}}\fi
\expandafter\ifx\csname urlprefix\endcsname\relax\def\urlprefix{URL: }\fi

\bibitem[{Andrews et~al.(2010)Andrews, Ganti, Haenggi, Jindal and Weber}]{A10}
Andrews, J.~G., Ganti, R.~K., Haenggi, M., Jindal, N. and Weber, S. (2010) A
  primer on spatial modeling and analysis in wireless networks.
\newblock \textit{Communications Magazine, IEEE}, \textbf{48}, 156--163.

\bibitem[{Ba and Coeurjolly(2020)}]{BC}
Ba, I. and Coeurjolly, J.-F. (2020) High-dimensional inference for
  inhomogeneous {G}ibbs point processes.

\bibitem[{Babu and Feigelson(1996)}]{BF96}
Babu, G.~J. and Feigelson, E.~D. (1996) Spatial point processes in astronomy.
\newblock \textit{Journal of Statistical Planning and Inference}, \textbf{50},
  311--326.

\bibitem[{Baccelli and B\l{}laszczyszyn(2009)}]{BB1}
Baccelli, F. and B\l{}laszczyszyn, B. (2009) \textit{Stochastic Geometry and
  Wireless Networks: Volume {I} Theory}.
\newblock Hanover, MA, USA: B.Now Publishers Inc.
\newblock \urlprefix\url{https://doi.org/10.1561/1300000006}.

\bibitem[{Baddeley et~al.(2014)Baddeley, Coeurjolly, Rubak and
  Waagepetersen}]{BCRW}
Baddeley, A., Coeurjolly, J.-F., Rubak, E. and Waagepetersen, R. (2014)
  Logistic regression for spatial {G}ibbs point processes.
\newblock \textit{Biometrika}, \textbf{101}, 377--392.

\bibitem[{Baddeley et~al.(2015)Baddeley, Rubak and Turner}]{spatstat}
Baddeley, A., Rubak, E. and Turner, R. (2015) \textit{Spatial Point Patterns:
  Methodology and Applications}.
\newblock London: Chapman and Hall/CRC Press.

\bibitem[{Baddeley et~al.(2000)Baddeley, Møller and
  Waagepetersen}]{Baddeley2000}
Baddeley, A.~J., Møller, J. and Waagepetersen, R. (2000) Non- and
  semi-parametric estimation of interaction in inhomogeneous point patterns.
\newblock \textit{Statistica Neerlandica}, \textbf{54}, 329--350.
\newblock
  \urlprefix\url{https://onlinelibrary.wiley.com/doi/abs/10.1111/1467-9574.00144}.

\bibitem[{Coeurjolly and Rubak(2013)}]{CR13}
Coeurjolly, J.-F. and Rubak, E. (2013) Fast covariance estimation for
  innovations computed from a spatial {G}ibbs point process.
\newblock \textit{Scandinavian Journal of Statistics}, \textbf{40}, 669--684.

\bibitem[{Condit(1998)}]{C}
Condit, R. (1998) \textit{Tropical Forest Census Plots}.
\newblock Berlin, Germany, and Georgetown, Texas: Springer-Verlag and R. G.
  Landes Company.

\bibitem[{Condit et~al.(2000)Condit, Ashton, Baker, Bunyavejchewin,
  Gunatilleke, Gunatilleke, Hubbell, Foster, Itoh, LaFrankie, Lee, Losos,
  Manokaran, Sukumar and Yamakura}]{Condit2000}
Condit, R., Ashton, P.~S., Baker, P., Bunyavejchewin, S., Gunatilleke, S.,
  Gunatilleke, N., Hubbell, S.~P., Foster, R.~B., Itoh, A., LaFrankie, J.~V.,
  Lee, H.~S., Losos, E., Manokaran, N., Sukumar, R. and Yamakura, T. (2000)
  Spatial patterns in the distribution of tropical tree species.
\newblock \textit{Science}, \textbf{288}, 1414--1418.
\newblock \urlprefix\url{https://science.sciencemag.org/content/288/5470/1414}.

\bibitem[{Condit et~al.(1999)Condit, Ashton, Manokaran, LaFrankie, Hubbell and
  Foster}]{CAMLHF}
Condit, R., Ashton, P.~S., Manokaran, N., LaFrankie, J.~V., Hubbell, S.~P. and
  Foster, R.~B. (1999) Dynamics of the forest communities at {P}asoh and
  {B}arro {C}olorado: comparing two 50-ha plots.
\newblock \textit{Philosophical transactions of the Royal Society of London.
  Series B, Biological sciences}, \textbf{354}, 1739--1748.

\bibitem[{Connor and Hill(1995)}]{CH95}
Connor, C.~B. and Hill, B.~E. (1995) Three nonhomogeneous {P}oisson models for
  the probability of basaltic volcanism: application to the {Y}ucca mountain
  region, {N}evada.
\newblock \textit{Journal of Geophysical Research: Solid Earth (1978-2012)},
  \textbf{100}, 10107--10125.

\bibitem[{Daley and Vere-Jones(2003)}]{DV}
Daley, D.~J. and Vere-Jones, D. (2003) \textit{An introduction to the theory of
  point processes}, vol.~1.
\newblock New York: Probability and its Applications. Springer-Verlag.

\bibitem[{Daley and Vere-Jones(2008)}]{DV2}
--- (2008) \textit{An introduction to the theory of point processes}, vol.~2.
\newblock New York: Probability and its Applications. Springer-Verlag.

\bibitem[{Daniel et~al.(2018)Daniel, Horrocks and Umphrey}]{DHU}
Daniel, J., Horrocks, J. and Umphrey, G.~J. (2018) Penalized composite
  likelihoods for inhomogeneous {G}ibbs point process models.
\newblock \textit{Computational Stat. and Data Analysis}, \textbf{124},
  104--116.

\bibitem[{Deyi et~al.(November 13, 2020)Deyi, Liu, Ye, Cadotte and He}]{Yin}
Deyi, Y., Liu, Y., Ye, Q., Cadotte, M. and He, F. (November 13, 2020) Trait
  dissimilarity and hierarchy predict spatial co-occurrence patterns of tree
  species in a subtropical forest.
\newblock \textit{Authorea}.

\bibitem[{Dixon(2002)}]{D}
Dixon, P.~M. (2002) Nearest-neighbor contingency table analysis of spatial
  segregation for several species.
\newblock \textit{Ecoscience}, \textbf{9}, 142--151.

\bibitem[{Fiksel(1988)}]{F}
Fiksel, T. (1988) Estimation of interaction potentials of {G}ibbsian point
  processes.
\newblock \textit{Statistics}, \textbf{19}, 77--86.

\bibitem[{Fl\"ugge et~al.(2014)Fl\"ugge, Olhede and Murrell}]{FOM}
Fl\"ugge, A.~J., Olhede, S.~C. and Murrell, D.~J. (2014) A method to detect
  subcommunities from multivariate spatial associations.
\newblock \textit{Methods in Ecology and Evolution}, \textbf{5}, 1214--1224.

\bibitem[{Geyer(1999)}]{G}
Geyer, C.~J. (1999) Likelihood inference for spatial point processes:
  Likelihood and computation.
\newblock In \textit{Stochastic Geometry: Likelihood and Computation} (eds.
  W.~Kendall, O.~Barndroff-Nielsen and M.~N. van Lieshout), 141--172. London:
  Chapman and Hall/CRC.

\bibitem[{Good and Whipple(1982)}]{GW}
Good, B.~J. and Whipple, S.~A. (1982) Tree spatial patterns: South {C}arolina
  bottomland and swamp forests.
\newblock \textit{Bulletin of the Torrey Botanical Club}, \textbf{109},
  529--536.

\bibitem[{Goulard et~al.(1996)Goulard, S\"arkk\"a and Grabarnik}]{GSG}
Goulard, M., S\"arkk\"a, A. and Grabarnik, P. (1996) Parameter estimation for
  marked {G}ibbs point processes through the maximum pseudolikelihood method.
\newblock \textit{Scandinavian Journal of Statistics}, \textbf{23}, 365--379.

\bibitem[{Hubbell et~al.(2012)Hubbell, Condit and Foster}]{HCF}
Hubbell, S., Condit, R. and Foster, R. (2012) {B}arro {C}olorado forest census
  plot data.
\newblock \url{http://ctfs.si.edu/webatlas/datasets/bci}.

\bibitem[{John et~al.(2007)John, Dalling, Harms, Yavitt, Stallard, Mirabello,
  Hubbell, Valencia, Navarrete, Vallejo and Foster}]{John2007PNAS_soil}
John, R., Dalling, J.~W., Harms, K.~E., Yavitt, J.~B., Stallard, R.~F.,
  Mirabello, M., Hubbell, S.~P., Valencia, R., Navarrete, H., Vallejo, M. and
  Foster, R.~B. (2007) Soil nutrients influence spatial distributions of
  tropical tree species.
\newblock \textit{Proceedings of the National Academy of Sciences},
  \textbf{104}, 864--869.
\newblock \urlprefix\url{https://www.pnas.org/content/104/3/864}.

\bibitem[{Jones et~al.(1994)Jones, Sharitz, James and Dixon}]{JSJD}
Jones, R.~H., Sharitz, R.~R., James, S.~M. and Dixon, P.~M. (1994) Tree
  population dynamics in seven {S}outh {C}arolina mixed-species forests.
\newblock \textit{Bulletin of the Torrey Botanical Club}, \textbf{121},
  360--368.

\bibitem[{Kallenberg(1983)}]{K}
Kallenberg, O. (1983) \textit{Random measures}.
\newblock Akademie-Verlag, 3 edn.

\bibitem[{Kupers et~al.(2019)Kupers, Wirth, Engelbrecht and R\"uger}]{KWER}
Kupers, S.~J., Wirth, C., Engelbrecht, B. M.~J. and R\"uger, N. (2019) Dry
  season soil water potential maps of a 50 hectare tropical forest plot on
  {B}arro {C}olorado {I}sland, {P}anama.
\newblock \textit{Scientific Data}, \textbf{6}, 63.

\bibitem[{Lombardo et~al.(2018)Lombardo, Opitz and Huser}]{lombardo}
Lombardo, L., Opitz, T. and Huser, R. (2018) Point process-based modeling of
  multiple debris flow landslides using inla: an application to the 2009
  messina disaster.
\newblock \textit{Stochastic Environmental Research and Risk Assessment},
  \textbf{32}, 2179--2198.

\bibitem[{Mohler et~al.(2011)Mohler, Short, Brantingham, Schoenberg and
  Tita}]{MSBST}
Mohler, G.~O., Short, M.~B., Brantingham, P.~J., Schoenberg, F.~P. and Tita,
  G.~E. (2011) Self-exciting point process modeling of crime.
\newblock \textit{Journal of the American Statistical Association},
  \textbf{106}, 100--108.

\bibitem[{M{\o}ller and Berthelsen(2012)}]{MB}
M{\o}ller, J. and Berthelsen, K.~K. (2012) Transforming spatial point processes
  into {P}oisson processes using random superposition.
\newblock \textit{Advances in Applied Probability}, \textbf{44}, 42--62.

\bibitem[{M{\o}ller and Waagepetersen(2004)}]{MW}
M{\o}ller, J. and Waagepetersen, R.~P. (2004) \textit{Statistical Inference and
  Simulation for Spatial Point Processes}.
\newblock Chapman and Hall.

\bibitem[{Nam and D'Agostino(2002)}]{auc}
Nam, B.-H. and D'Agostino, R.~B. (2002) \textit{Discrimination Index, the Area
  Under the ROC Curve}, 267--279.
\newblock Boston, MA: Birkh{\"a}user Boston.
\newblock \urlprefix\url{https://doi.org/10.1007/978-1-4612-0103-8_20}.

\bibitem[{Ovaskainen et~al.(2017)Ovaskainen, Tikhonov, Norberg,
  Guillaume~Blanchet, Duan, Dunson, Roslin and Abrego}]{ovaskainen2017make}
Ovaskainen, O., Tikhonov, G., Norberg, A., Guillaume~Blanchet, F., Duan, L.,
  Dunson, D., Roslin, T. and Abrego, N. (2017) How to make more out of
  community data? a conceptual framework and its implementation as models and
  software.
\newblock \textit{Ecology Letters}, \textbf{20}, 561--576.

\bibitem[{Penttinen et~al.(1992)Penttinen, Stoyanell and Henttonen}]{PSH}
Penttinen, A., Stoyanell, D. and Henttonen, H.~M. (1992) Marked point processes
  in forest statistics.
\newblock \textit{Forest Science}, \textbf{638}, 806--824.

\bibitem[{Punchi‐Manage et~al.(2013)Punchi‐Manage, Getzin, Wiegand,
  Kanagaraj, Gunatilleke, Gunatilleke, Wiegand and Huth}]{PGWKGGWH}
Punchi‐Manage, R., Getzin, S., Wiegand, T., Kanagaraj, R., Gunatilleke, C.
  V.~S., Gunatilleke, I. A. U.~N., Wiegand, K. and Huth, A. (2013) Effects of
  topography on structuring local species assemblages in a {S}ri {L}ankan mixed
  dipterocarp forest.
\newblock \textit{Journal of Ecology}, \textbf{101}, 149--160.

\bibitem[{Rajala et~al.(2018)Rajala, Murrell and Olhede}]{RMO}
Rajala, T., Murrell, D.~J. and Olhede, S.~C. (2018) Detecting multivariate
  interactions in spatial point patterns with {G}ibbs models and variable
  selection.
\newblock \textit{Journal of the Royal Statistical Society: Series C},
  \textbf{67}, 1237--1273.

\bibitem[{Seabloom et~al.(2005)Seabloom, Bjørnstad, Bolker and
  Reichman}]{seabloom}
Seabloom, E.~W., Bjørnstad, O.~N., Bolker, B.~M. and Reichman, O.~J. (2005)
  Spatial signature of environmental heterogeneity, dispersal, and competition
  in successional grasslands.
\newblock \textit{Ecological Monographs}, \textbf{75}, 199--214.
\newblock
  \urlprefix\url{https://esajournals.onlinelibrary.wiley.com/doi/abs/10.1890/03-0841}.

\bibitem[{Seri et~al.(2015)Seri, Shtilerman and Shnerb}]{seri}
Seri, E., Shtilerman, E. and Shnerb, N.~M. (2015) The glocal forest.
\newblock \textit{PLOS ONE}, \textbf{10}, 1--9.
\newblock \urlprefix\url{https://doi.org/10.1371/journal.pone.0126117}.

\bibitem[{Shen et~al.(2013)Shen, He, Waagepetersen, Sun, Hao, Chen and
  Yu}]{shen2013quantifying}
Shen, G., He, F., Waagepetersen, R., Sun, I.-F., Hao, Z., Chen, Z.-S. and Yu,
  M. (2013) Quantifying effects of habitat heterogeneity and other clustering
  processes on spatial distributions of tree species.
\newblock \textit{Ecology}, \textbf{94}, 2436--2443.

\bibitem[{Thompson(1955)}]{TH55}
Thompson, H. (1955) Spatial point processes, with applications to ecology.
\newblock \textit{Biometrika}, \textbf{42}, 102--115.

\bibitem[{Uriarte et~al.(2004{\natexlab{a}})Uriarte, Condit, Canham and
  Hubbell}]{UCCH}
Uriarte, M., Condit, R., Canham, C.~D. and Hubbell, S.~P. (2004{\natexlab{a}})
  A spatially explicit model of sapling growth in a tropical forest: Does the
  identity of neighbours matter?
\newblock \textit{Journal of Ecology}, \textbf{92}, 348--360.

\bibitem[{Uriarte et~al.(2004{\natexlab{b}})Uriarte, Condit, Canham and
  Hubbell}]{uriarte2004spatially}
--- (2004{\natexlab{b}}) A spatially explicit model of sapling growth in a
  tropical forest: does the identity of neighbours matter?
\newblock \textit{Journal of Ecology}, \textbf{92}, 348--360.

\bibitem[{Waagepetersen et~al.(2016)Waagepetersen, Guan, Jalilian and
  Mateu}]{WGJM}
Waagepetersen, R., Guan, Y., Jalilian, A. and Mateu, J. (2016) Analysis of
  multispecies point patterns by using multivariate log-{G}aussian {C}ox
  processes.
\newblock \textit{Journal of the Royal Statistical Society: Series C},
  \textbf{65}, 77--96.

\bibitem[{Waller and Gotway(2004)}]{WG04}
Waller, L.~A. and Gotway, C.~A. (2004) \textit{Applied Spatial Statistics for
  Public Health Data}.
\newblock John Wiley \& Sons.

\bibitem[{Weiher et~al.(2011)Weiher, Freund, Bunton, Stefanski, Lee and
  Bentivenga}]{weiher2011advances}
Weiher, E., Freund, D., Bunton, T., Stefanski, A., Lee, T. and Bentivenga, S.
  (2011) Advances, challenges and a developing synthesis of ecological
  community assembly theory.
\newblock \textit{Philosophical Transactions of the Royal Society B: Biological
  Sciences}, \textbf{366}, 2403--2413.

\bibitem[{Wiegand et~al.(2007)Wiegand, Gunatilleke and
  Gunatilleke}]{wiegand2007species}
Wiegand, T., Gunatilleke, S. and Gunatilleke, N. (2007) Species associations in
  a heterogeneous sri lankan dipterocarp forest.
\newblock \textit{The American Naturalist}, \textbf{170}, E77--E95.

\end{thebibliography}


\begin{thebibliography}{3}
\expandafter\ifx\csname natexlab\endcsname\relax\def\natexlab#1{#1}\fi
\expandafter\ifx\csname url\endcsname\relax
  \def\url#1{\texttt{#1}}\fi
\expandafter\ifx\csname urlprefix\endcsname\relax\def\urlprefix{URL: }\fi

\bibitem[{Ambler and Silverman(2009)}]{AZ}
Ambler, G.~K. and Silverman, B.~W. (2009) Perfect simulation of spatial point
  processes using dominated coupling from the past with application to a
  multiscale area-interaction point process.

\bibitem[{Berthelsen and M{\o}ller(2002)}]{BM}
Berthelsen, K.~K. and M{\o}ller, J. (2002) A primer on perfect simulation for
  spatial point processes.
\newblock \textit{Bulletin of the Brazilian Mathematical Society}, \textbf{33},
  351--367.

\bibitem[{M{\o}ller and Waagepetersen(2004)}]{MW}
M{\o}ller, J. and Waagepetersen, R.~P. (2004) \textit{Statistical Inference and
  Simulation for Spatial Point Processes}.
\newblock Chapman and Hall.

\end{thebibliography}
\end{document}


\maketitle
	
	\section{Computation of the Papangelou conditional intensity}
	
	The Papangelou conditional intensity can be computed from the density $j$ by using the formula $\pi((x,i,m),\omega):=j(\omega\cup\{(x,i,m)\})/j(\omega)$ for $(x,i,m)\notin\omega$.
	Indeed, we have
	\begin{align*}
	&\pi((x,i,m),\omega)\\
	&\quad:=\frac{j(\omega\cup\{(x,i,m)\})}{j(\omega)}\\
	&\quad=\exp\Biggl[\beta_{i,0}+\sum_{k=1}^K\beta_{i,k}X_k(x)\\
	&\quad\qquad+\sum_{i_2=1}^p\biggl(\sum_{z=(x_1,i_1,m_1)\in\omega\cup\{(x,i,m)\}}\alpha_{i_1,i_2}u(z,((\omega\cup\{(x,i,m)\})\setminus\{z\})_{i_2})\\
	&\quad\qquad\qquad\qquad\qquad\qquad\qquad\qquad\qquad\qquad-\sum_{z=(x_1,i_1,m_1)\in\omega}\alpha_{i_1,i_2}u(z,(\omega\setminus\{z\})_{i_2})\biggr)\\
	&\quad\qquad+\sum_{i_2=1}^p\biggl(\sum_{z=(x_1,i_1,m_1)\in\omega\cup\{(x,i,m)\}}\gamma_{i_1,i_2}v(z,((\omega\cup\{(x,i,m)\})\setminus\{z\})_{i_2})\\
	&\quad\qquad\qquad\qquad\qquad\qquad\qquad\qquad\qquad\qquad-\sum_{z=(x_1,i_1,m_1)\in\omega}\gamma_{i_1,i_2}v(z,(\omega\setminus\{z\})_{i_2})\biggr)\Biggr]\\
	&\quad=\exp\Biggl[\beta_{i,0}+\sum_{k=1}^K\beta_{i,k}X_k(x)\\
	&\quad\qquad+\sum_{i_2=1}^p\biggl(\alpha_{i,i_2}u((x,i,m),\omega_{i_2})\\
	&\quad\qquad\quad+\sum_{z=(x_1,i_1,m_1)\in\omega}\alpha_{i_1,i_2}\bigl(u(z,((\omega\cup\{(x,i,m)\})\setminus\{z\})_{i_2})-u(z,(\omega\setminus\{z\})_{i_2})\bigr)\biggr)\\
	&\quad\qquad+\sum_{i_2=1}^p\biggl(\gamma_{i,i_2}v((x,i,m),\omega_{i_2})\\
	&\quad\qquad\quad+\sum_{z=(x_1,i_1,m_1)\in\omega}\gamma_{i_1,i_2}\bigl(v(z,((\omega\cup\{(x,i,m)\})\setminus\{z\})_{i_2})-v(z,(\omega\setminus\{z\})_{i_2})\bigr)\biggr)\Biggr].
	\end{align*}
	Swapping the roles of $i_1$ and $i_2$ in two of the summations and using the symmetry of $\alpha$ and $\gamma$ yields
	\begin{align*}
	&\pi((x,i,m),\omega)\\
	&\quad=\exp\Biggl[\beta_{i,0}+\sum_{k=1}^K\beta_{i,k}X_k(x)\\
	&\quad\quad+\sum_{i_2=1}^p\alpha_{i,i_2}u((x,i,m),\omega_{i_2})\\
	&\quad\quad\quad+\sum_{i_1=1}^p\sum_{z=(x_2,i_2,m_2)\in\omega}\alpha_{i_1,i_2}\bigl(u(z,((\omega\cup\{(x,i,m)\})\setminus\{z\})_{i_1})-u(z,(\omega\setminus\{z\})_{i_1})\bigr)\\
	&\quad\quad+\sum_{i_2=1}^p\gamma_{i,i_2}v((x,i,m),\omega_{i_2})\\
	&\quad\quad\quad+\sum_{i_1=1}^p\sum_{z=(x_2,i_2,m_2)\in\omega}\gamma_{i_1,i_2}\bigl(v(z,((\omega\cup\{(x,i,m)\})\setminus\{z\})_{i_1})-v(z,(\omega\setminus\{z\})_{i_1})\bigr)\Biggr].
	\end{align*}
	It remains to notice that $u(z,((\omega\cup\{(x,i,m)\})\setminus\{z\})_{i_1})=u(z,(\omega\setminus\{z\})_{i_1})$ and $v(z,((\omega\cup\{(x,i,m)\})\setminus\{z\})_{i_1})=v(z,(\omega\setminus\{z\})_{i_1})$ when $i_1\neq i$, to obtain 
	\begin{align}
	&\pi((x,i,m),\omega)\nonumber\\
	&\quad=\exp\Biggl[\beta_{i,0}+\sum_{k=1}^K\beta_{i,k}X_k(x)
	+\sum_{i_2=1}^p\alpha_{i,i_2}u((x,i,m),\omega_{i_2})\nonumber\\
	&\quad\quad\quad+\sum_{z=(x_2,i_2,m_2)\in\omega}\alpha_{i,i_2}\bigl[u(z,((\omega\setminus\{z\})\cup\{(x,i,m)\})_{i})-u(z,(\omega\setminus\{z\})_{i})\bigr]\nonumber\\
	&\quad\quad+\sum_{i_2=1}^p\gamma_{i,i_2}v((x,i,m),\omega_{i_2})\nonumber\\
	&\quad\quad\quad+\sum_{z=(x_2,i_2,m_2)\in\omega}\gamma_{i,i_2}\bigl[v(z,((\omega\setminus\{z\})\cup\{(x,i,m)\})_{i})-v(z,(\omega\setminus\{z\})_{i})\bigr]\Biggr].\label{eq:ModelPapangelou}
	\end{align}
	
	\section{Conditional distribution of species conditional on others}
	
	In this section, we state and prove two new results making explicit the conditional distribution of some species conditional on others, under additional hypotheses.
	This is used in the main text in our model assessment of the Barro Colorado Island case study.
	In the following proposition, we derive conditions which ensure that one species conditional on others is an inhomogeneous Poisson point process.
	
    \begin{prop}
    \label{prop:ConditionalIntensity}
        Let $i_0\in\{1,\dots,p\}$ be a given species.
        We denote the saturated pairwise interaction Gibbs point process by $\xi$, the point process consisting in all individuals of species $i_0$ by $\xi_{i_0}:=\{(x,i_0,m)\in\xi\}$ and those not of species $i_0$ by $\xi_{-i_0}:=\{(x,i,m)\in\xi\;:\;i\neq i_0\}$.
        Assume that $N=\infty$ and that there are no intra-species interactions for species $i_0$, i.e., $\alpha_{i_0,i_0}=\gamma_{i_0,i_0}=0$.
        Then $(\xi_{i_0}\mid \xi_{-i_0})$ is an inhomogeneous Poisson point process with intensity proportional to $\pi((x,i_0,m),\xi_{-i_0})$, i.e., the Papangelou conditional intensity computed at a point of species $i_0$, conditional on the configuration $\xi_{-i_0}$.
    \end{prop}
    \begin{proof}
        We assume without loss of generality that $i_0=1$.
	The density of the saturated pairwise interaction Gibbs point process at a configuration $\omega\in\mathcal N$ is denoted by abuse of notation by $j(\xi_1,\xi_{-1})$, where $\xi_1:=\{(x,1,m)\in\xi\}$ and $\xi_{-1}:=\{(x,i,m)\in\xi\;:\;i\neq 1\}$.
	We note that $j(\xi_1,\xi_{-1})>0$ $\mathbb P$-almost surely.
    We start by computing the density of $(\xi_1\mid\xi_{-1})$.
    Let $h$ and $G$ be measurable non-negative functions, and $\zeta_1$ and $\zeta_{-1}$ be two independent unit-rate Poisson processes defined respectively on $W\times\{1\}\times\mathbb R$ and $W\times\{2,\dots,p\}\times\mathbb R$.
    We have
    \begin{align*}
    &\mathbb E\biggl[\exp\Bigl(-\sum_{x\in\xi_1}h(x)\Bigr)G(\xi_{-1})\biggr]\\
    &\qquad=\mathbb E\biggl[\exp\Bigl(-\sum_{x\in \zeta_1}h(x)\Bigr)G(\zeta_{-1})j(\zeta_1,\zeta_{-1})\biggr]\\
    &\qquad=\mathbb E\biggl[G(\zeta_{-1})\mathbb E\biggl[\exp\Bigl(-\sum_{x\in \zeta_1}h(x)\Bigr)j(\zeta_1,\zeta_{-1})\;\Big|\;\zeta_{-1}\biggr]\biggr]\\
    &\qquad=\mathbb E\biggl[G(\zeta_{-1})\int\exp\Bigl(-\sum_{x\in \eta}h(x)\Bigr)j(\eta,\zeta_{-1})\,\mathbb P_{\zeta_1}(\mathrm d\eta)\biggr]\\
    &\qquad=\mathbb E\biggl[G(\xi_{-1})\frac{1}{j(\xi_1,\xi_{-1})}\int\exp\Bigl(-\sum_{x\in \eta}h(x)\Bigr)j(\eta,\xi_{-1})\,\mathbb P_{\zeta_1}(\mathrm d\eta)\biggr]\\
    &\qquad=\mathbb E\biggl[G(\xi_{-1})\mathbb E\biggl[\frac{1}{j(\xi_1,\xi_{-1})}\;\Big|\;\xi_{-1}\biggr]\int\exp\Bigl(-\sum_{x\in \eta}h(x)\Bigr)j(\eta,\xi_{-1})\,\mathbb P_{\zeta_1}(\mathrm d\eta)\biggr],
    \end{align*}
    where $\mathbb P_{\zeta_1}$ denotes the distribution of $\zeta_1$.
    We deduce from this that
    \begin{equation*}
    \mathbb E\biggl[\exp\Bigl(-\sum_{x\in\xi_1}h(x)\Bigr)\;\Big|\;\xi_{-1}\biggr]
    =\mathbb E\biggl[\frac{1}{j(\xi_1,\xi_{-1})}\;\Big|\;\xi_{-1}\biggr]
    \mathbb E\biggl[\exp\Bigl(-\sum_{x\in\zeta}h(x)\Bigr)j(\zeta,\xi_{-1})\biggr],
    \end{equation*}
    where $\zeta$ is a unit-rate Poisson process defined on $W\times\{1\}\times\mathbb R$ and independent of $\xi_{-1}$.
    Therefore, the density of $(\xi_1\mid\xi_{-1})$ with respect to $\zeta_1$ is
    \begin{equation}
    \label{eq:ConditionalPDF}
    j_1(\omega_1):=Cj(\omega_1,\xi_{-1}),
    \end{equation}
    where $C>0$ is a proportionality constant that depends on $\xi_{-1}$.
    
    Assume now that $N=\infty$, $\alpha_{1,1}=\gamma_{1,1}=0$.
    In the manuscript we consider two possibilities for $u$ and $v$, which we call unmarked and marked cases.
    Here we restrict ourselves to the unmarked case, since the proof is essentially unchanged in the marked case.
    We also assume that $\gamma_{i,j}=0$ for all $i,j$ in order to simplify the presentation, noting indeed that the medium-range interactions only add extra terms analogous to those corresponding to the short-range interactions in the density.
    By \eqref{eq:ConditionalPDF}, the density of $(\xi_1\mid\xi_{-1})$ is proportional to $j(\omega_1,\xi_{-1})$ which can be computed as
    \begin{align*}
	&j(\omega_1,\xi_{-1})\\
	&\qquad=C\exp\Biggl[\sum_{(x,1,m)\in\omega_1}\biggl(\beta_{1,0}+\sum_{k=1}^K\beta_{1,k}X_k(x)\biggr)
	+\sum_{(x,i,m)\in\xi_{-1}}\biggl(\beta_{i,0}+\sum_{k=1}^K\beta_{i,k}X_k(x)\biggr)\\
	&\qquad\qquad\qquad+2\sum_{i_2=2}^p\sum_{(x_1,1,m_1)\in\omega_1}\sum_{(x_2,i_2,m_2)\in\xi_{-1}}\alpha_{1,i_2}\varphi_{R^{\mathrm S}_{i_2,1}}(\|x_1-x_2\|)\\
	&\qquad\qquad\qquad+\sum_{i_1,i_2=2}^p\sum_{z=(x_1,i_1,m_1)\in\xi_{-1}}\alpha_{i_1,i_2}\sum_{(x_2,i_2,m_2)\in(\xi_{-1}\setminus\{z\})_{i_2}}\varphi_{R^{\mathrm S}_{i_1,i_2}}(\|x_1-x_2\|)\Biggr]\\
	&\qquad=F(\xi_{-1})\times\prod_{(x,1,m)\in\omega_1}\exp\Biggl[\biggl(\beta_{1,0}+\sum_{k=1}^K\beta_{1,k}X_k(x)\biggr)\\
	&\qquad\qquad\qquad\qquad\qquad\qquad+2\sum_{i_2=2}^p\sum_{(x_2,i_2,m_2)\in\xi_{-1}}\alpha_{1,i_2}\varphi_{R^{\mathrm S}_{i_2,1}}(\|x-x_2\|)\Biggr],
	\end{align*}
	where $F(\xi_{-1})$ collects all terms involving points in $\xi_{-1}$.
	This is indeed the density of an inhomogeneous Poisson point process with intensity
	\begin{equation*}
	   \lambda_1(x,m)
	   =\exp\Biggl[\beta_{1,0}+\sum_{k=1}^K\beta_{1,k}X_k(x)\\
    	+2\sum_{i_2=2}^p\sum_{(x_2,i_2,m_2)\in\xi_{-1}}\alpha_{1,i_2}\varphi_{R^{\mathrm S}_{i_2,1}}(\|x-x_2\|)\Biggr],
	\end{equation*}
	and it is readily checked that this quantity is proportional to $j(\{(x,1,m)\},\xi_{-1})$, and therefore to $\pi((x,1,m),\xi_{-1})$.
    \end{proof}
    The following result shows that $\alpha_{i,j}$ and $\gamma_{i,j}$, $i\neq j$, control the correlation between species $i$ and $j$.
    \begin{prop}
    \label{prop:ConditionalIntensityTwoSpecies}
        Let $i_0,i_1\in\{1,\dots,p\}$ be two given species, and let the notation of Proposition~\ref{prop:ConditionalIntensity} prevail.
        We denote by $\xi_{-\{i_0,i_1\}}:=\{(x,i,m)\in\xi\;:\;i\not\in\{i_0,i_1\}\}$.
        If $\alpha_{i_0,i_1}=\gamma_{i_0,i_1}=0$, then $\xi_{i_0}$ and $\xi_{i_1}$ are independent conditionally on $\xi_{-\{i_0,i_1\}}$.
    \end{prop}
    \begin{proof}
        We assume without loss of generality that $i_0=1$ and $i_1=2$.
	We also assume that $\alpha_{1,2}=0$ and as in the proof of Proposition~1, we assume that $\gamma_{i,j}=0$ for all $i,j$ in order to simplify the presentation.
	By the same argument as in the proof of Proposition~1, the density of $((\xi_1,\xi_2)\mid\xi_{-\{1,2\}})$ with respect to unit rate Poisson point processes is
    \begin{equation*}
    j_{1,2}(\omega_1,\omega_2):=Cj(\omega_1,\omega_2,\xi_{-\{1,2\}}),
    \end{equation*}
    for a constant $C>0$.
	We have
	\begin{align*}
	&j(\omega_1,\omega_2,\xi_{-\{1,2\}})\\
	&\qquad=C\exp\Biggl[
	\sum_{(x,1,m)\in\omega_1}\biggl(\beta_{1,0}+\sum_{k=1}^K\beta_{1,k}X_k(x)\biggr)
	+\sum_{(x,2,m)\in\omega_2}\biggl(\beta_{2,0}+\sum_{k=1}^K\beta_{2,k}X_k(x)\biggr)\\
	&\qquad\qquad\qquad\qquad\qquad\qquad+\sum_{(x,i,m)\in\xi_{-\{1,2\}}}\biggl(\beta_{i,0}+\sum_{k=1}^K\beta_{i,k}X_k(x)\biggr)\\
	&\qquad\qquad\qquad\qquad\qquad\qquad+\sum_{z=(x_1,1,m_1)\in\omega_1}\alpha_{1,1}u(z,\omega_1\setminus\{z\})\\
	&\qquad\qquad\qquad\qquad\qquad\qquad+\sum_{i_2=3}^p\sum_{z=(x_1,1,m_1)\in\omega_1}\alpha_{1,i_2}u(z,(\xi_{-\{1,2\}})_{i_2})\\
	&\qquad\qquad\qquad\qquad\qquad\qquad+\sum_{z=(x_1,2,m_1)\in\omega_2}\alpha_{2,2}u(z,\omega_2\setminus\{z\})\\
	&\qquad\qquad\qquad\qquad\qquad\qquad+\sum_{i_2=3}^p\sum_{z=(x_1,2,m_1)\in\omega_2}\alpha_{2,i_2}u(z,(\xi_{-\{1,2\}})_{i_2})\\
	&\qquad\qquad\qquad\qquad\qquad\qquad+\sum_{i_1=3}^p\sum_{z=(x_1,i_1,m_1)\in\xi_{-\{1,2\}}}\alpha_{i_1,1}u(z,\omega_1)\\
	&\qquad\qquad\qquad\qquad\qquad\qquad+\sum_{i_1=3}^p\sum_{z=(x_1,i_1,m_1)\in\xi_{-\{1,2\}}}\alpha_{i_1,2}u(z,\omega_2)\\
	&\qquad\qquad\qquad\qquad\qquad\qquad+\sum_{i_1,i_2=3}^p\sum_{z=(x_1,i_1,m_1)\in\xi_{-\{1,2\}}}\alpha_{i_1,i_2}u(z,(\xi_{-\{1,2\}}\setminus\{z\})_{i_2})
	\Biggr]\\
	&\qquad\qquad\qquad\qquad\qquad\qquad=F(\xi_{-\{1,2\}})\times F_1(\omega_1,\xi_{-\{1,2\}})\times F_2(\omega_2,\xi_{-\{1,2\}}),
	\end{align*}
	where $F_i(\omega_i,\xi_{-\{1,2\}})$ collects all terms involving points in $\omega_i$, $i=1,2$ and $F(\xi_{-\{1,2\}})$ collects all terms involving points in $\xi_{-\{1,2\}}$ only.
	This implies that $\xi_1$ and $\xi_2$ are independent conditionally on $\xi_{-\{1,2\}}$.
    \end{proof}
    
    \section{Details of the coupling from the past algorithm}
	
	A naive implementation of the `coupling from the past' algorithm based on Section~11.2.6 of \cite{MW} does not suffice. 
	Indeed, computing $\alpha_{\text{min}}$ and $\alpha_{\text{max}}$ on p.~230 of \cite{MW} is extremely inefficient, since in our case the point process is neither attractive nor repulsive, cf. equation (6.7) again in \cite{MW}.
	In another reference \cite{AZ}, the authors consider a version of the coupling from the past algorithm which does not require the point process to be attractive nor repulsive, however our model does not satisfy their hypotheses on p.~5.
	
	\begin{sloppypar}
		Instead, we follow the original idea on p.~361 of \cite{BM}, namely we decompose the Papangelou conditional intensity computed above as $\pi(z,\omega)=\pi_{\text{increasing}}(z,\omega)\cdot\pi_{\text{decreasing}}(z,\omega)$, where $\pi_{\text{increasing}}$ is obtained by replacing $\alpha_{i,i_2}$ with $\max(\alpha_{i,i_2},0)$ and setting $\gamma_{i,i_2}=0$, in the formula, and $\pi_{\text{decreasing}}((x,i,m),\omega):=\exp\bigl[A((x,i,m),\omega)+G((x,i,m),\omega)\bigr]$, for 
		\begin{multline*}
		A((x,i,m),\omega)
		:=\sum_{i_2=1}^p\min(\alpha_{i,i_2},0)\biggl(u((x,i,m),\omega_{i_2})\\
		+\sum_{z=(x_2,i_2,m_2)\in\omega}\bigl[u(z,((\omega\setminus\{z\})\cup\{(x,i,m)\})_i)-u(z,(\omega\setminus\{z\})_i)\bigr]\biggr),
		\end{multline*}
		and
		\begin{multline*}
		G((x,i,m),\omega)
		:=\sum_{i_2=1}^p\min(\gamma_{i,i_2},0)\biggl(v((x,i,m),\omega_{i_2})\\
		+\sum_{z=(x_2,i_2,m_2)\in\omega}\bigl[v(z,((\omega\setminus\{z\})\cup\{(x,i,m)\})_i)-v(z,(\omega\setminus\{z\})_i)\bigr]\biggr).
		\end{multline*}
	\end{sloppypar}
	We say that a measurable function $f:\mathbb S\times\mathcal N\to\mathbb R$ is \textit{increasing} if for $z\in\mathbb{S}$, $f(z,\omega)\le f(z,\eta)$ for $\omega\subset\eta$ and that it is \textit{decreasing} if for $z\in\mathbb{S}$, $f(z,\omega)\ge f(z,\eta)$ for $\omega\subset\eta$.
	We assume hereon that $u$ and $v$ are increasing, and note that this assumption holds for $u_{\text{unmarked}}$, $u_{\text{marked}}$, $v_{\text{unmarked}}$ and $v_{\text{marked}}$ from the main text.
	
	It is readily checked that $\pi_{\text{increasing}}$ is increasing and $\pi_{\text{decreasing}}$ is decreasing.
	It can also be seen that $\pi_{\text{decreasing}}$ is bounded by $1$ and $\pi_{\text{increasing}}$ is bounded by $h_1$ defined in the main text.
	
	Thus, returning to the argument laid out on p.~361 of \cite{BM}, we can apply the coupling from the past algorithm with $\alpha_{\text{min}}(z,l,u)=\pi_{\text{increasing}}(z,u)\cdot\pi_{\text{decreasing}}(z,l)$ and $\alpha_{\text{max}}(z,l,u)=\pi_{\text{increasing}}(z,l)\cdot\pi_{\text{decreasing}}(z,u)$.
	The complete algorithm including the particular details of how the quantities above are used may be found on p.~360 of \cite{BM} or in Section~11.2.6 of \cite{MW}.
	
	\section{Local stability lemma}
	
	\begin{lemma}
		\label{lem:AihuaLemma}
		Let $N\ge1$, and let $\varphi:[0,\infty)\to[0,1]$ be a bounded decreasing function.
		Letting
		\begin{equation*}
		u((x,i,m),\omega):=\max_{\eta\subset\omega\ :\ |\eta|\le N}\sum_{(x_1,i_1,m_1)\in\eta}\varphi(\|x-x_1\|),
		\end{equation*}
		we have
		\begin{equation}
		\label{eq:BoundToProve}
		0\le\sum_{z\in\omega}\bigl[u(z,(\omega\setminus\{z\})\cup\{(x,i,m)\})-u(z,\omega\setminus\{z\})\bigr]\le5N,
		\end{equation}
		for every $(x,i,m)\in\mathbb S$ and $\omega\in\mathcal N$.
		In addition, the bound \eqref{eq:BoundToProve} is attained for some $(x,i,m)$, $\omega$ and $\varphi$.
	\end{lemma}
	\begin{proof}
		In the proof, we shall fix $(x,i,m)\in\mathbb S$ and $\omega\in\mathcal N$, and show that 
		\begin{equation*}
		S_N:=\bigl\{z\in\omega\ :\ u(z,(\omega\setminus\{z\})\cup\{(x,i,m)\})\neq u(z,\omega\setminus\{z\})\bigr\}
		\end{equation*}
		has no more than $5N$ elements.
		This, combined with the fact that $\varphi$ is bounded by one, ensures that \eqref{eq:BoundToProve} holds.
		
		The geometric interpretation of our proof is as follows. 
		Take $z_1=(x_1,i_1,m_1)$ with $x_1$ as a furthest point in $S_N$ from $x$, collect all points in $S_N$ with locations that are closer to $x_1$ than $x_1$ is to $x$ to form $V_{z_1}$, repeat the procedure for $S_N\backslash V_{z_1}$ to get $z_2$ and $V_{z_2}$, and so on, until all points in $S_N$ are exhausted.
		Then, show that each $V_{z_i}$ contains no more than $N$ points and there are no more than five $V_{z_i}$'s. 
		More precisely, we let  $V_{z_0}:=\emptyset$ and construct $V_{z_l}$ for $l\ge1$ recursively as follows.
		We set 
		\begin{equation*}
		A_l:=S_N\setminus\bigl(V_{z_0}\cup\cdots\cup V_{z_{l-1}}\bigr),
		\end{equation*}
		and if $A_l\neq\emptyset$, choose $z_l=(x_l,i_l,m_l)\in A_l$ such that
		\begin{equation}
		\label{eq:zlCharacterization}
		\varphi(\|x-x_l\|)=\min_{(x',i',m')\in A_l}\varphi(\|x-x'\|),
		\end{equation}
		and define
		\begin{equation*}
		V_{z_l}:=\bigl\{(x',i',m')\in A_l\ :\ \varphi(\|x-x_l\|)\le\varphi(\|x'-x_l\|)\bigr\}.
		\end{equation*}
		Since the number of points in $\omega$ is finite, the procedure above has a finite number of steps, and we have
		\begin{equation*}
		S_N=\bigcup_{i=0}^kV_{z_i},
		\end{equation*}
		for some finite $k\ge0$.
		
		Note that there are at most $N$ elements in $V_{z_l}$, for any $l\in\{0,\dots,k\}$.
		Indeed, for $l\ge1$, if $|\omega|>N$ let us write
		\begin{equation*}
		u(z_l,\omega\setminus\{z_l\})=\sum_{i=1}^N\varphi(\|x_l-y_i\|),
		\end{equation*}
		for some $(y_1,i_1,m_1),\dots,(y_N,i_N,m_N)\in\omega$
		(if $|\omega|\le N$ then the result is trivially true).
		Since $z_l\in S_N$, we have that $\varphi(\|x-x_l\|)>\varphi(\|y_i-x_l\|)$ for at least one of the $y_i$'s.
		This implies in particular that 
		\begin{equation}
		\label{eq:Inequalityzlyi}
		\varphi(\|x-x_l\|)\le\varphi(\|y_i-x_l\|),
		\end{equation} 
		for at most $N-1$ of the $y_i$'s.
		Inequality \eqref{eq:Inequalityzlyi} also holds with $x_l$ in place of $y_i$ since $\varphi$ is decreasing, and so it holds for at most $N$ points of $\omega$.
		
		In the following, it only remains to prove that $k\le5$.
		Let $i,j$ be such that $1\le i<j\le k$, and write $z_i=(x_i,i_i,m_i)$ and $z_j=(x_j,i_j,m_j)$.
		By construction, $z_j\notin V_{z_i}$, and so 
		\begin{equation}
		\label{eq:FirstInequalityVarphi}
		\varphi(\|x-x_i\|)>\varphi(\|x_i-x_j\|),
		\end{equation}
		and since $i<j$, by \eqref{eq:zlCharacterization} we have
		\begin{equation}
		\label{eq:SecondInequalityVarphi}
		\varphi(\|x-x_i\|)\le\varphi(\|x-x_j\|).
		\end{equation}
		From \eqref{eq:FirstInequalityVarphi} and the fact that $\varphi$ is decreasing, we obtain $\widehat{x_ixx_j}>\widehat{x_ix_jx}$ where $\widehat{abc}$ is the angle formed by $ab$ and $bc$.
		Similarly, from \eqref{eq:SecondInequalityVarphi} and the fact that $\varphi$ is decreasing, we get $\widehat{x_ix_jx}\ge\widehat{x_jx_ix}$.
		Thus, $\widehat{x_ixx_j}$ is strictly larger than $60^\circ$, and so $k<6$.
		This concludes the proof of the first part of the statement.
		
		Let us now fix $(x,i,m)\in\mathbb S$ and choose $\varphi=\mathbf1_{[0,1]}$.
		Then, one can take $\omega_0\in\mathcal N$ with $5N$ points within a distance $1$ of $x$, such that each of the five groups of $N$ points is within $0.01$ of one of the five vertices of a regular pentagon circumscribed in the circle around $x$ of radius $1$.
		We illustrate this choice of $\omega_0$ in Figure~\ref{fig:ConfigurationAttainingBound}.
		It is easy to check that for any $z\in\omega_0$, $u(z,\omega_0\setminus\{z\})=N-1$ and $u(z,(\omega_0\setminus\{z\})\cup\{(x,i,m)\})=N$.
		Since $\omega_0$ has $5N$ points, the bound \eqref{eq:BoundToProve} is attained.
		
		\definecolor{ffqqqq}{rgb}{1,0,0}
		\definecolor{ududff}{rgb}{0.30196078431372547,0.30196078431372547,1}
		\definecolor{uuuuuu}{rgb}{0.26666666666666666,0.26666666666666666,0.26666666666666666}
		\definecolor{zzttqq}{rgb}{0.6,0.2,0}
		\definecolor{cqcqcq}{rgb}{0.7529411764705882,0.7529411764705882,0.7529411764705882}
		\begin{figure}
			\centering
			\scalebox{0.3}{
				\begin{tikzpicture}[line cap=round,line join=round,>=triangle 45,x=1cm,y=1cm]
				\clip(-18.119212979566868,-8.979723897621295) rectangle (18.80935588981407,21.314426118277755);
				\fill[line width=2pt,color=zzttqq,fill=zzttqq,fill opacity=0.10000000149011612] (-8,-5) -- (8,-5) -- (12.94427190999916,10.216904260722451) -- (0,19.621468297402025) -- (-12.944271909999157,10.216904260722458) -- cycle;
				\draw [line width=2pt,color=zzttqq] (-8,-5)-- (8,-5);
				\draw [line width=2pt,color=zzttqq] (8,-5)-- (12.94427190999916,10.216904260722451);
				\draw [line width=2pt,color=zzttqq] (12.94427190999916,10.216904260722451)-- (0,19.621468297402025);
				\draw [line width=2pt,color=zzttqq] (0,19.621468297402025)-- (-12.944271909999157,10.216904260722458);
				\draw [line width=2pt,color=zzttqq] (-12.944271909999157,10.216904260722458)-- (-8,-5);
				\draw [line width=2pt] (0,6.011055363769388) circle (13.610412933632636cm);
				\draw [shift={(8,-5)},line width=2pt]  plot[domain=0.6773000523873031:3.71644145640657,variable=\t]({1*1.3327843891551772*cos(\t r)+0*1.3327843891551772*sin(\t r)},{0*1.3327843891551772*cos(\t r)+1*1.3327843891551772*sin(\t r)});
				\draw [shift={(-8,-5)},line width=2pt]  plot[domain=-0.5792302459171781:2.465760381419458,variable=\t]({1*1.3356870766217148*cos(\t r)+0*1.3356870766217148*sin(\t r)},{0*1.3356870766217148*cos(\t r)+1*1.3356870766217148*sin(\t r)});
				\draw [shift={(-12.944271909999157,10.21690426072246)},line width=2pt]  plot[domain=-1.8365701460833126:1.1919521568528755,variable=\t]({1*1.3165779436006202*cos(\t r)+0*1.3165779436006202*sin(\t r)},{0*1.3165779436006202*cos(\t r)+1*1.3165779436006202*sin(\t r)});
				\draw [shift={(0,19.621468297402025)},line width=2pt]  plot[domain=3.1899531665321526:6.232968587188152,variable=\t]({1*1.3159000371654668*cos(\t r)+0*1.3159000371654668*sin(\t r)},{0*1.3159000371654668*cos(\t r)+1*1.3159000371654668*sin(\t r)});
				\draw [shift={(12.94427190999916,10.216904260722453)},line width=2pt]  plot[domain=1.9328511309604555:4.961607808274138,variable=\t]({1*1.3032577131920384*cos(\t r)+0*1.3032577131920384*sin(\t r)},{0*1.3032577131920384*cos(\t r)+1*1.3032577131920384*sin(\t r)});
				\draw [->,line width=2pt,color=ffqqqq] (0,19.621468297402025) -- (-1.314361563533217,19.55785549901539);
				\draw [->,line width=2pt,color=ffqqqq] (0,6.011055363769388) -- (13.149611043981228,2.4994930880861777);
				\begin{huge}
				\draw [fill=uuuuuu] (0,6.011055363769388) circle (4pt);
				\draw[color=uuuuuu] (0.23904193006795776,6.606616285558765) node {$x$};
				\draw [fill=ududff] (-0.6634115866666477,18.78276961904762) circle (3.5pt);
				\draw [fill=ududff] (0.19552708000001934,18.78276961904762) circle (3.5pt);
				\draw [fill=ududff] (-12.47381825333332,10.224059333333336) circle (3.5pt);
				\draw [fill=ududff] (-12.903287586666654,9.273091523809526) circle (3.5pt);
				\draw [fill=ududff] (-8.485888729523793,-4.347221619047611) circle (3.5pt);
				\draw [fill=ududff] (-7.657626443809508,-4.684661809523801) circle (3.5pt);
				\draw [fill=ududff] (7.557858508571451,-4.4085743809523725) circle (3.5pt);
				\draw [fill=ududff] (7.0363600323809745,-5.29818942857142) circle (3.5pt);
				\draw [fill=ududff] (12.28202117523812,9.487826190476193) circle (3.5pt);
				\draw [fill=ududff] (12.005933746666692,10.254735714285717) circle (3.5pt);
				\draw[color=ffqqqq] (0.29963023009975587,20.299572092745137) node {$\epsilon=0.01$};
				\draw[color=ffqqqq] (6.721990033470354,4.9) node {$R=1$};
				\draw [fill=ududff] (-0.0033112700592346197,19.284718067212516) circle (3.5pt);
				\draw [fill=ududff] (0.9963956804654339,19.224129767180717) circle (3.5pt);
				\draw [fill=ududff] (12.386996086443476,11.105297562919773) circle (3.5pt);
				\draw [fill=ududff] (12.90199663671376,9.408825162029427) circle (3.5pt);
				\draw [fill=ududff] (7.024931533629345,-4.677954595363631) circle (3.5pt);
				\draw [fill=ududff] (8.4487565843766,-4.314424795172842) circle (3.5pt);
				\draw [fill=ududff] (-7.001259923731915,-5.253543445665713) circle (3.5pt);
				\draw [fill=ududff] (-8,-4) circle (3.5pt);
				\draw [fill=ududff] (-12.423912776577845,9.257354411949931) circle (3.5pt);
				\draw [fill=ududff] (-12,10) circle (3.5pt);
				\end{huge}
				\end{tikzpicture}
			}
			\caption{Points in the configuration $\omega_0$ (as blue dots) in the case $N=4$.} \label{fig:ConfigurationAttainingBound}
		\end{figure}
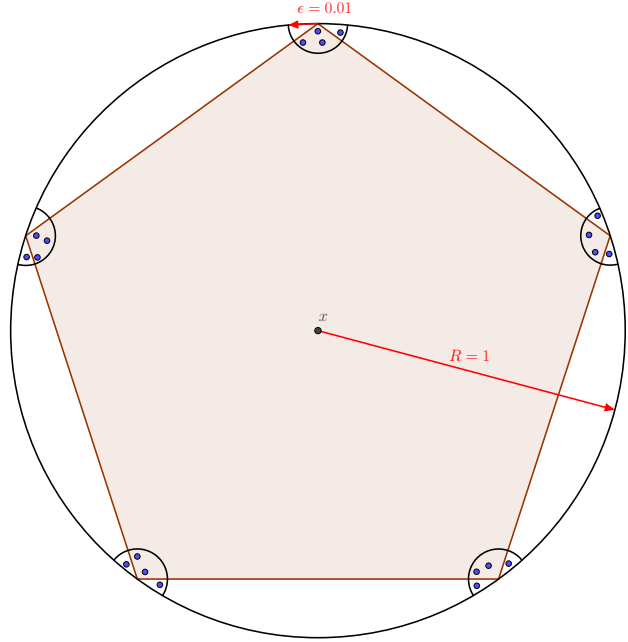
	\end{proof}
		
	\section{Simulation study with seven species}
	
	Compared to the simulation in the main text, this study involves significantly more interacting species.
	We consider a `saturated pairwise interaction Gibbs point process' on the square region $W=[0,1]^2$, consisting of $p=7$ species, with no marks, and whose distribution is driven by a single geospatial covariate, $X_1(x,y)=x-0.5$.
	We consider uniform short-range interaction radii of $R^{\mathrm S}=0.03$, medium-range interaction radii of $R^{\mathrm M}=0.04$ and long-range interaction radii of $R^{\mathrm L}=0.08$.
	The coefficient corresponding to the covariate $X_1$ is 
	\begin{equation*}
	    \beta_1^T=(-0.50, -0.33, -0.17, 0.00, 0.17,0.33, 0.50),
	\end{equation*}
	and the interaction matrices are given by
	\begin{equation*}
	\alpha=\begin{pmatrix}
-0.4 & 0 & -0.1 & 0 & 0 & -0.2 & 0.2\\
0 & -0.1 & 0.2 & -0.1 & -0.1 & -0.2 & 0.2\\
-0.1 & 0.2 & 0.2 & 0.2 & -0.2 & -0.2 & 0.2\\
0 & -0.1 & 0.2 & -0.4 & 0.2 & -0.1 & -0.1\\
0 & -0.1 & -0.2 & 0.2 & -0.3 & -0.2 & 0.1\\
-0.2 & -0.2 & -0.2 & -0.1 & -0.2 & 0 & -0.2\\
0.2 & 0.2 & 0.2 & -0.1 & 0.1 & -0.2 & 0.2
\end{pmatrix}
	\end{equation*}
	and 
	\begin{equation*}
	\gamma=\begin{pmatrix}
0 & 0.1 & 0.2 & -0.2 & -0.1 & 0 & -0.2\\
0.1 & 0.2 & 0 & -0.1 & 0 & -0.2 & -0.2\\
0.2 & 0 & 0.1 & 0.1 & 0 & -0.2 & 0.2\\
-0.2 & -0.1 & 0.1 & 0.1 & 0 & -0.2 & 0.2\\
-0.1 & 0 & 0 & 0 & -0.4 & 0.2 & -0.1\\
0 & -0.2 & -0.2 & -0.2 & 0.2 & 0 & -0.1\\
-0.2 & -0.2 & 0.2 & 0.2 & -0.1 & -0.1 & -0.4
\end{pmatrix}.
	\end{equation*}
	
	\begin{figure}[!ht]
		\centering
		\includegraphics[trim={0cm 0cm 0cm 1.19cm},clip,width=0.95\textwidth]{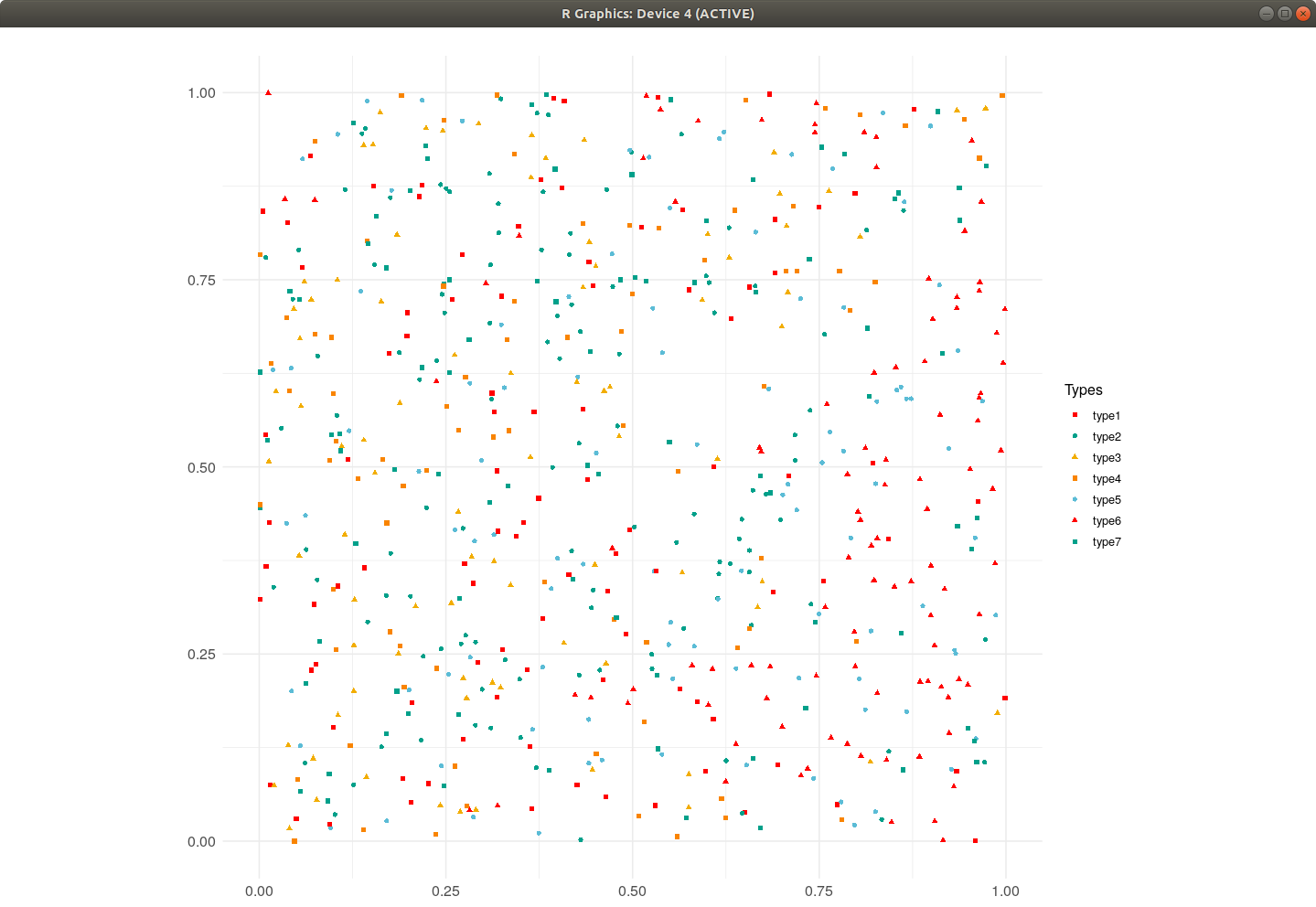}
		\caption{\label{fig:additional_typical_sample} Typical sample considered in this numerical experiment.}
	\end{figure}
	
	We set the saturation parameter $N$ to 2, take as the short-range potential the exponential, and choose the normal medium-range potential.
	With so many interacting species, the final distribution of individuals is highly sensitive to the choice of the intercept vector $\beta_0$.
	We chose the intercept vector by a step procedure which ensured that samples had around $100$ individuals in each species. 
	Our procedure yielded $\beta_0^T=(5.28, 4.85, 3.36, 4.94, 5.73, 6.02, 4.95)$.
	In order to illustrate our experiment, we plot in Figure~\ref{fig:additional_typical_sample} a typical sample from this point process.
	
	\begin{figure}[!ht]
		\centering
		\includegraphics[trim={0cm 0cm 0cm 0cm},clip,width=0.95\textwidth]{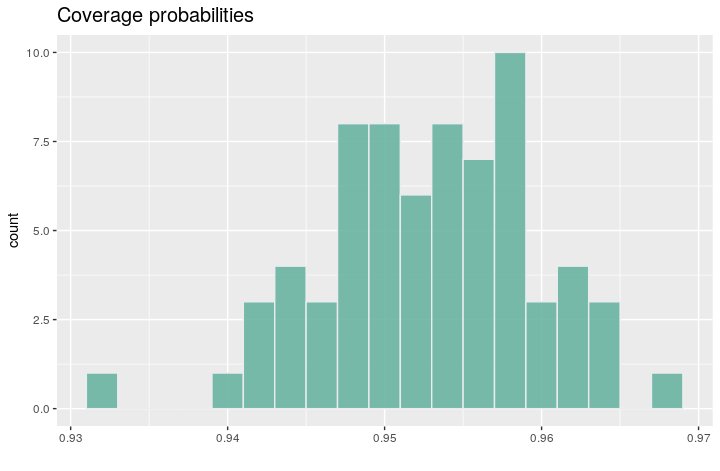}
		\caption{\label{fig:coverage_prob} Average coverage probabilities.}
	\end{figure}
	
	\begin{figure}[!ht]
		\centering
		\includegraphics[trim={0 0 0 0},clip,width=0.9\textwidth]{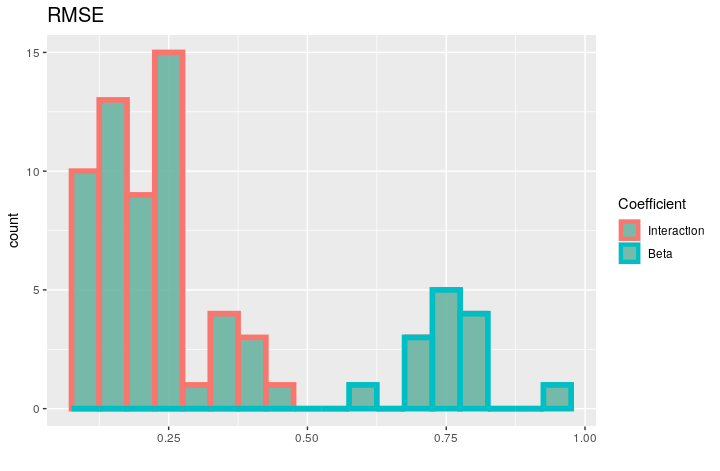}
		\caption{\label{fig:average_rmse} Average RMSE for the estimates of $\alpha$ and $\gamma$ (red border) and of $\beta_0$ and $\beta_1$ (blue border).}
	\end{figure}
	
	We sampled $1,000$ independent draws of this point process with $1,000,000$ steps of the Metropolis-Hastings algorithm.
	The saturation parameter, interaction distances, and interaction shapes were set to their true values.
	We then fit each of the samples by the logistic regression introduced in the main text, and produced asymptotic confidence intervals.
	Since the number of parameters is quite large, we present the results graphically in Figures~\ref{fig:coverage_prob} and \ref{fig:average_rmse}.
	The results are quite convincing, and the model is not showing any sign of misperformance when studying seven species jointly.
	
	\section{R Package}
	\label{sec:package}
		
	The model described in this manuscript has been implemented in the R language (R Core Team, 2019).
	The package is available on IF's personal Github at \url{https://github.com/iflint1/ppjsdm}.
	The numerical experiments described in the main text and in the previous section can be reproduced by executing the scripts included with the above package.
	
	\bibliography{library}